\documentclass[%
 aip,
 amsmath,amssymb,
reprint,%
]{revtex4-2}

\usepackage{algorithmic}
\usepackage{graphicx}
\usepackage{textcomp}
\usepackage{xcolor}
\usepackage[caption=false,font=footnotesize]{subfig}
\usepackage{orcidlink}
\usepackage{float}  
\usepackage{braket}  
\usepackage{soul}
\usepackage{dcolumn}
\usepackage{bm}
\usepackage[utf8]{inputenc}
\usepackage[T1]{fontenc}
\usepackage{mathptmx}
\usepackage{etoolbox}
\usepackage{fancyhdr}

\makeatletter
\def\@email#1#2{%
 \endgroup
 \patchcmd{\titleblock@produce}
  {\frontmatter@RRAPformat}
  {\frontmatter@RRAPformat{\produce@RRAP{*#1\href{mailto:#2}{#2}}}\frontmatter@RRAPformat}
  {}{}
}%
\makeatother

\begin{document}

\preprint{AIP/123-QED}

\title[]{Variational Quantum Eigensolver Models of Molecular Quantum Dot Cellular Automata}
\author{Nischal Binod Gautam}
 \affiliation{Department of Electrical and Computer Engineering, Baylor University.}
 	\email{nischal\_gautam1@baylor.edu}
\author{Enrique P. Blair}%
 \affiliation{Department of Electrical and Computer Engineering, Baylor University.}

\date{\today}

\begin{abstract}
  Molecular quantum-dot Cellular Automata (QCA) may provide low-power, high-speed computational hardware for processing classical information. Simulation and modeling play an important role in the design of QCA circuits because fully-coherent models of QCA scale exponentially with the number of devices, and such models are severely limited in size. For larger circuits, approximations become necessary. In the era of fault-tolerant quantum computation, however, it may become possible to model large QCA circuits without such limitations. Presently, this work explores the use of the noisy-intermediate scale quantum (NISQ) variational quantum eigensolver (VQE) method for estimating the ground state of QCA circuits. This is relevant because the computational result of a QCA calculation is encoded in the circuit's ground state. In this study, VQE is used to model logic circuits, including binary wires, 
  inverters, and majority gates. VQE models are performed ideal simulators, noisy simulators, and actual quantum hardware. This study demonstrates that VQE may indeed be used to model molecular QCA circuits. It is observed that using modern NISQ hardware, results are still quite sensitive to noise, so measures should be taken to minimize noise. These include simplifying the \(ansatz\) circuit whenever possible, and using low-noise hardware.
\end{abstract}

\maketitle

\begin{center}
\small
\textit{The following article has been submitted to \textbf{Journal of Applied Physics}. After it is published, it will be found at \href{https://publishing.aip.org/resources/librarians/products/journals}{link}.}
\end{center}

\section{Introduction}

Energy-efficient and low-power classical computing are especially desirable as rapid growth in AI usage drives  significant growth in global compute power consumption. A molecular implementation of quantum-dot cellular automata (QCA) \cite{lent1993QCA,orlov1997realization,snider1999quantum,lent2002Science-molecular} could provide low-power, high-speed , general-purpose computing with room temperature operation with nanometer-scale devices. \cite{tougaw1996dynamic,blair2016electric}

In anticipation of technologies that can arrange QCA molecules on a substrate with the requisite precision for information processing circuits, the simulation and modeling of molecular QCA 
circuits plays an important role in developing and demonstrating \textit{in silico} methods for low-power information processing and the write-in/readout of bits at the molecular scale. \cite{lent1997device,hennessy2001clocking,timler2002power,pulimeno2011-write-in,blair2019electric-field-input,cong2024circuits} Since classical models of such circuits scale exponentially with the number of molecular devices in a circuit, only a handful of molecular devices are tractable when arbitrary quantum correlations between devices are to be retained. This scaling of classical models raises the question whether quantum computational methods may be suitable for modeling QCA circuits. While quantum annealing methods have been previously explored\cite{retallick2014embedding}, variational quantum eigensolver (VQE) methods \cite{peruzzo2014variational} have not yet been published for modeling QCA circuits.

In this paper we explore the application of VQE methods to the modeling of molecular QCA circuits in both simulation and on actual hardware. Section \ref{sec:Background} provides a brief review of QCA and its molecular implementation. Section \ref{sec:Model} describes the method used here to model molecular QCA circuits using VQE. Results for basic molecular circuits are presented in Section \ref{sec:Results}. It is demonstrated here that VQE methods may be used to model QCA circuits on modern noisy-intermediate scale quantum (NISQ) hardware; however, results remain quite sensitive to noise. When possible, it is helpful to minimize noise by simplifying ansatz circuits to minimize the number of single-qubit and two-qubit operations.
 
\section{Background} \label{sec:Background}

\subsection{Development History}

\begin{figure}[htbp]
    \centering
    \subfloat[]{%
      \includegraphics[height=1.875in]{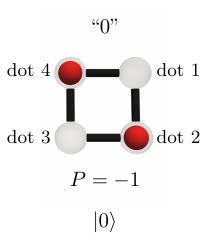}%
      \label{fig:subfig-0}%
    }\hfil
    \subfloat[]{%
      \includegraphics[height=1.875in]{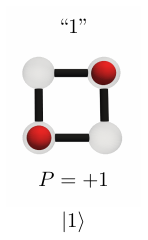}%
      \label{fig:subfig-1}%
    }
    \caption{A single bit of information is encoded by two different localized electronic charge 
    configurations of electrons (solid red spheres) on a system of four quantum dots (translucent grey spheres). Device switching occurs through charge tunneling, via paths indicated as black bars. Bit zero ``0'' [subfigure (a)] and bit ``1'' [subfigure (b)] are assigned the polarizations \(P = \mp 1\), respectively, and are respectively assigned quantum states \(\ket{0}\) and \(\ket{1}\). }
    \label{fig:1}
\end{figure}

In quantum-dot cellular automata, the configuration of mobile charge on a system of quantum dots is used to encode a binary state. This is illustrated in Figure \ref{fig:1}, where two mobile electrons on a system of four coupled dots provides two charge-localized states, which we identify as ``0'' and ``1,'' assigning them labels \(\ket{0}\) and \(\ket{1}\), respectively. We also assign these states a polarization of \(P = \mp 1\), where the polarization is a function of the mobile charge on each dot:
\begin{equation}
P = \frac{(\rho_1 + \rho_3) - (\rho_2 + \rho_4)}{\rho_1 + \rho_2 + \rho_3 + \rho_4} ,
\end{equation}
where \(\rho_{k}\) is the mobile charge on the \(k\)-th dot. This device is called a ``cell'' in QCA, and charge tunneling between the dots enables device switching.\cite{lent1993lines}

\begin{figure}[htbp]
    \centering
    \subfloat[]{%
      \includegraphics[width=0.21\textwidth]{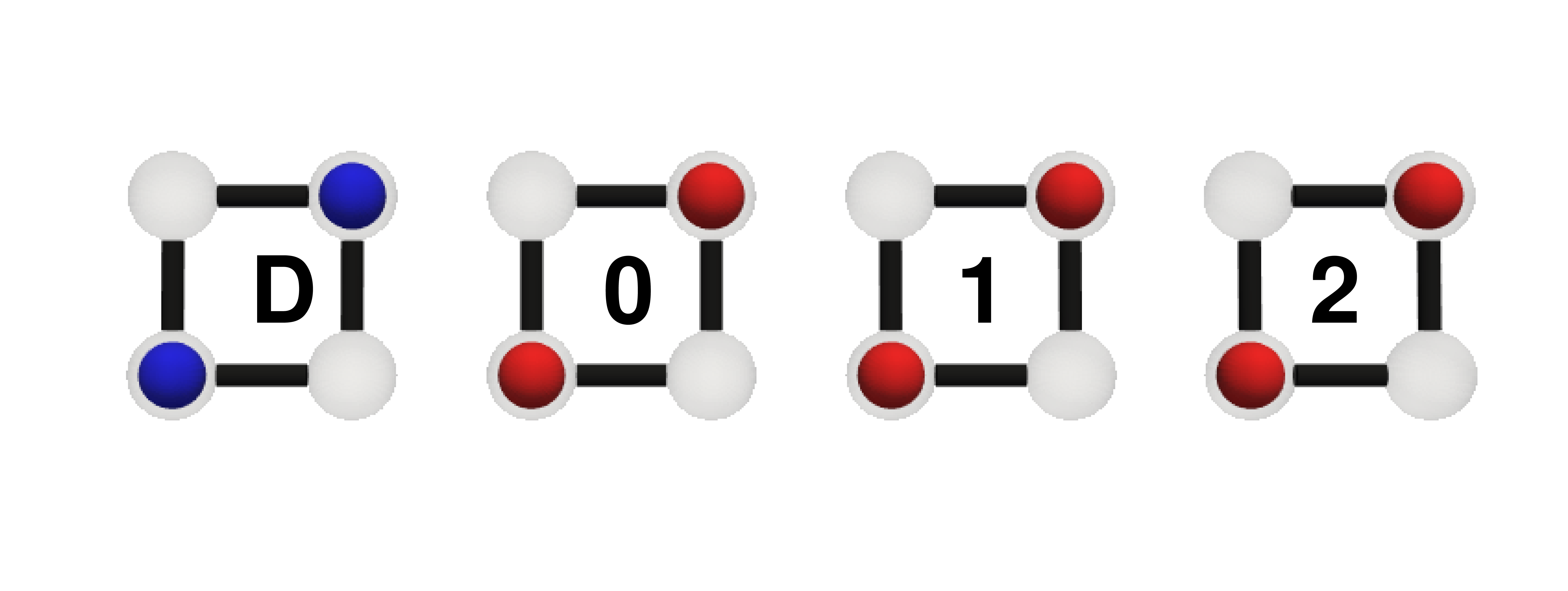}%
      \label{fig:1-3-cell}%
    }
    \subfloat[]{%
      \includegraphics[width=0.22\textwidth]{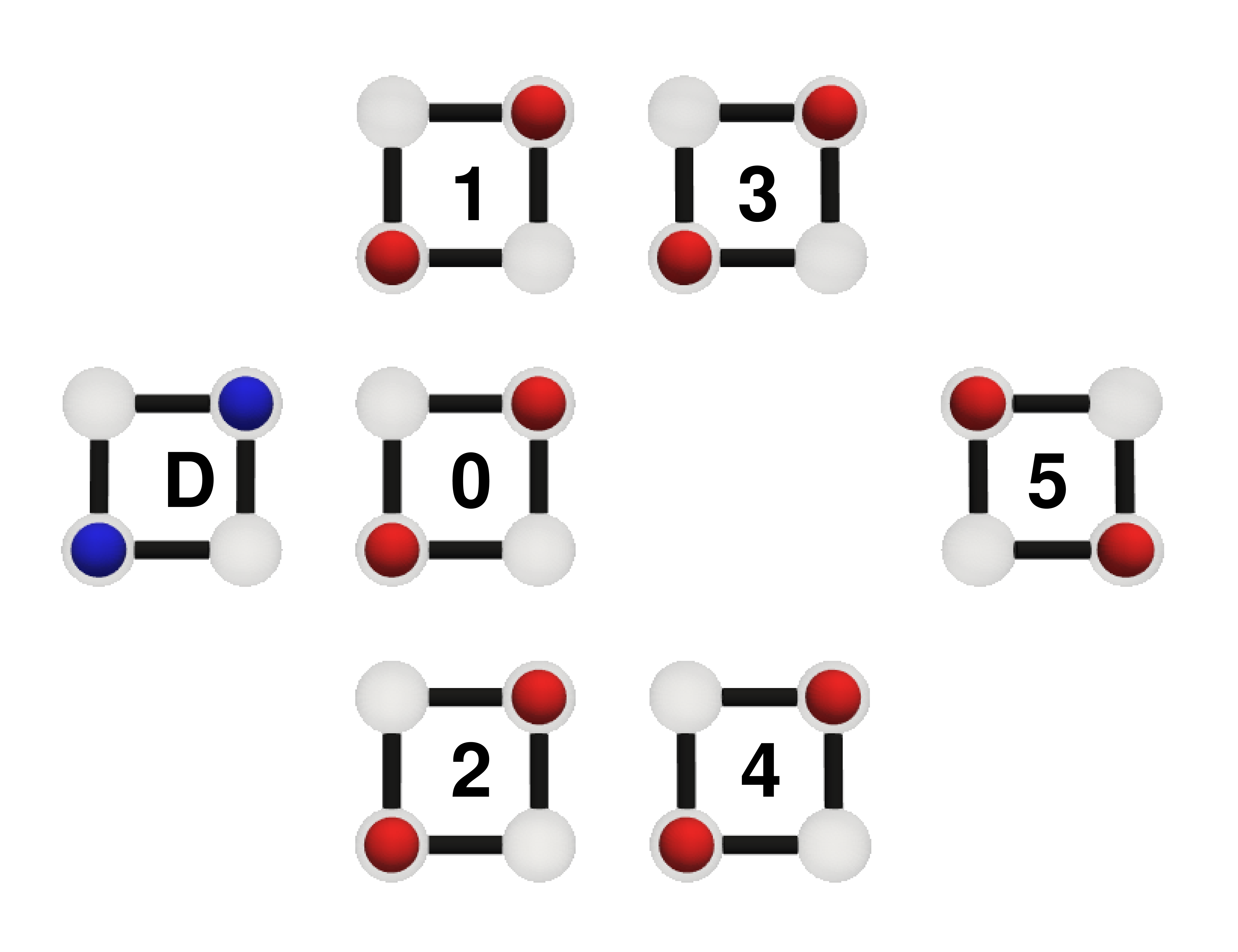}%
      \label{fig:2-6-inv}%
    }\\
    \subfloat[]{%
      \includegraphics[width=0.28\textwidth]{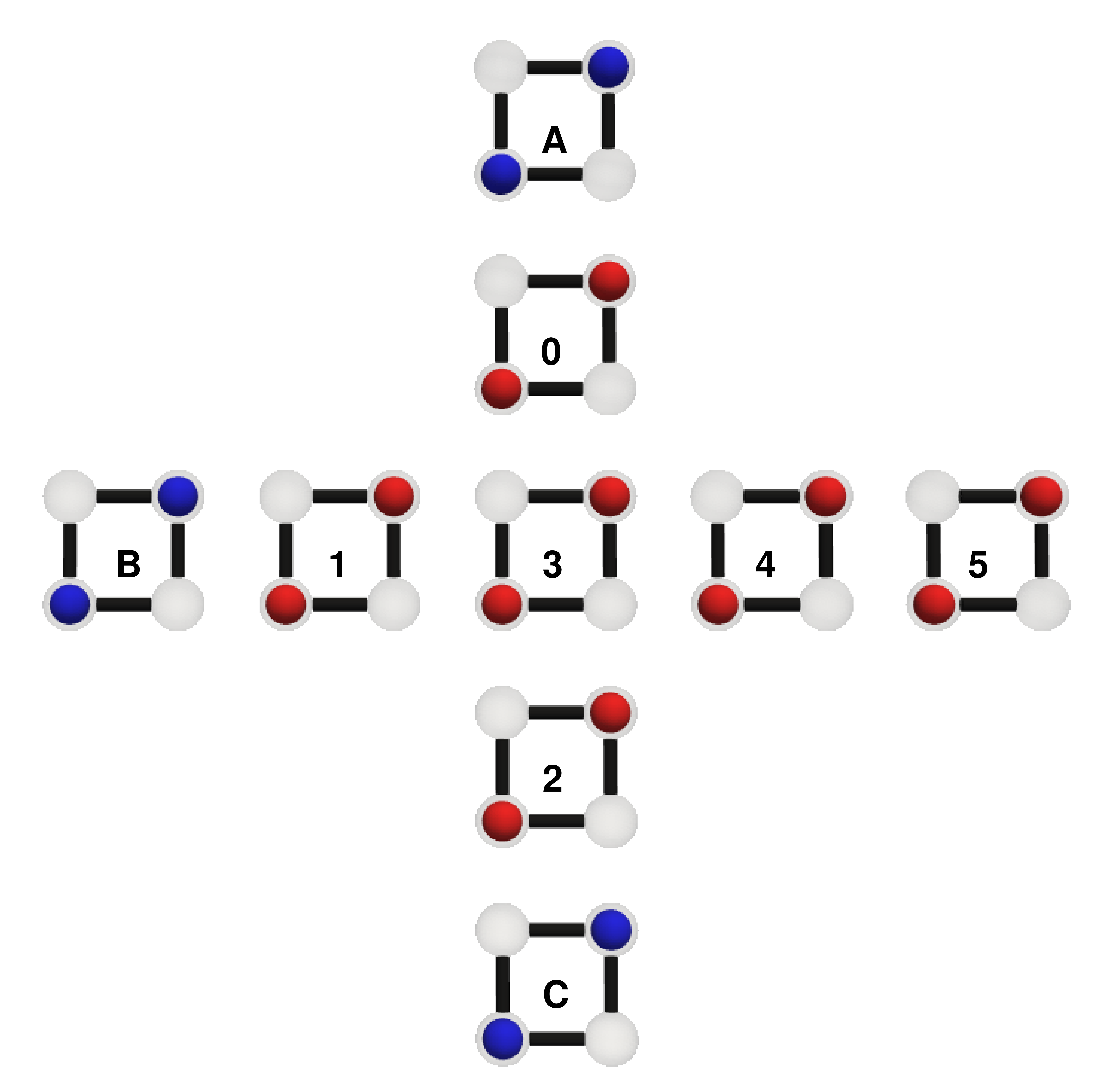}%
      \label{fig:majority1-logic}%
    }
    \caption{A logically complete set of circuits is possible in QCA. (a) The binary wire provides a path for data transmission. (b) The inverter and (c) majority gate support boolean logic operations. The inverter flips the input 
    bit using diagonal coupling, and the majority gate functions as a 
    programmable two-input AND/OR gate.}
    \label{fig:full-width-fig}
\end{figure}

Cells may be arranged on a surface to form circuits, in which cells couple locally through Coulomb interactions to enable information processing. A logically complete set of circuits is shown in Figure \ref{fig:full-width-fig}. In the binary wire  of Figure \ref{fig:1-3-cell}, an input bit specified on the driver cell, \(D\) (shown with blue mobile electrons for illustration), is copied down the line through nearest-neighbor interactions to cells 0, 1, and 2. This serves as the basic interconnective link in QCA circuits. The inverter of Figure \ref{fig:2-6-inv} fans out the bit from cell \(D\) into two copies on cells 3 and 4. The next-nearest-neighbor interaction with these bit copies causes a bit inversion on cell 5. Finally, the majority gate of Figure \ref{fig:majority1-logic} provides a natural logic gate in QCA. It accepts three input bits, \(A\), \(B\), and \(C\), and outputs the bit in the majority, implementing the function \(M(A, B, C) = AB + AC + BC\). Any one of the inputs may be leveraged as a control bit to configure the gate to function as either an AND or OR gate between the remaining inputs. For example, setting \(C=0\) yields \(M(A, B, 0) = AB\); and setting \(C=1\) yields \(M(A, B, 1) = A + B\). Designers have combined QCA logic into larger, more complex circuits, such as adders,\cite{wang2003-QCA-adder} multipliers,\cite{cho2009adder} and even entire processors.\cite{niemier1999logic}

\begin{figure}[htbp]
    \centering
    \subfloat[]{%
      \includegraphics[height=1.35in]{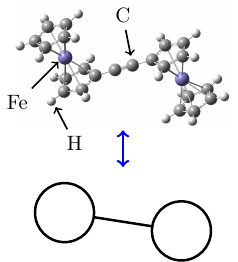}%
      \label{fig:dqd}%
    }\qquad
    \subfloat[]{%
      \includegraphics[height=1.4in]{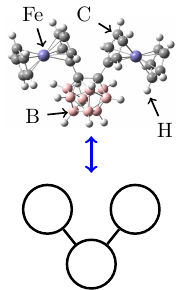}%
      \label{fig:tqd}%
    } \qquad
     \subfloat[]{%
      \includegraphics[height=1.4in]{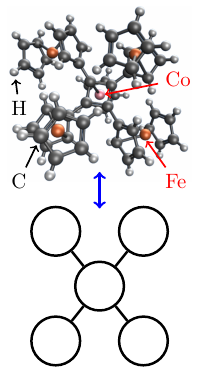}%
      \label{fig:qqd}%
    }
    \caption{Molecules may provide systems of coupled quantum dots. (a) Two iron centers (atoms colored purple) provide a coupled pair of molecular quantum dots.\cite{quardokus2013adsorption}
 (b) Two iron centers and a carborane cage provide a three-dot QCA system.\cite{christie2015synthesis} (c) Four iron centers provide corner dots, and a central Co atom provides a fifth 
dot that provides an ET path between the corner dots.\cite{jiao2005properties} In this paper we model molecules with direct tunneling between the corner dots and have no fifth central dot (see Figure \ref{fig:1}).}
    \label{fig:molecules}
\end{figure}

QCA cells have been implemented using systems of metal dots,\cite{orlov1997realization} semiconductor dots,\cite{macucci2003qca,gardelis2003realization} and at the \(\sim 1\)-nm scale using 
dangling bonds on a Si surface.\cite{haider2009controlled} Graphene quantum dots \cite{wang2011nanopatterned} have been considered as building blocks for QCA. In molecular QCA,\cite{lent2002Science-molecular,lent2003molecular,lent2003clocked,qi2003molecular} a mixed-valence molecule can function as either a whole cell or part of a cell. Here, redox centers on the molecule provide non-bonding orbitals that localize mobile charge. Some examples of this are shown in Figure \ref{fig:molecules}. In the molecule of Figure \ref{fig:dqd}, two Fe centers provide a pair of coupled molecular quantum dots, also known as a double-quantum-dot (DQD) system. A pair of DQDs could be used to implement a four-dot cell. Figure \ref{fig:tqd} shows two Fe centers and a carborane cage, which provides a net neutral, zwitterionic three-dot system \cite{christie2015synthesis} that could 
support clocked QCA operation.\cite{toth1999quasiadiabatic,orlov2000experimental-APL,kummamuru2001power,hennessy2001clocking,lent2003clocked,orlov2003clocked} A pair of these could be used to implement a clocked six-dot cell. A five-dot cell is shown in Figure \ref{fig:qqd}, in which the four corner dots provide the same two states as in the case of the four-dot cells illustrated earlier, and the central dot provides the 
tunneling path between all corner dots.\cite{jiao2005properties}

In this work, we focus on molecular QCA because their \(\sim 1\)-nm scale lends itself to bit energies robust at room temperature, extremely high device densities, 
and molecular devices with high intrinsic switching speeds.\cite{blair2016electric} Additionally, we focus on circuits built from the four-dot cells introduced earlier. Molecular 
QCA candidates are an ongoing subject of exploration through design, modeling, synthesis, and characterization.\cite{li2003molecular,lu2007molecular,lu2013counterion,christie2015synthesis,tsukerblat2018mQCA,graziano2019characterisation,liza2021asymmetric,liza2024ab,montenegro2023exploring} While clocking is important in QCA because it allows the latching of bits, power gain for the restoration of weakened bits, synchronized operation of circuits, and reduced power dissipation below bit energies through reversible 
erasures,\cite{timler2002power,kummamuru2002power,orlov2002two} we focus here on unclocked QCA cells because of their simplicity. 

In QCA, the ground state of a molecular circuit is important because it encodes the computational result of the circuit. Modeling the ground state of QCA circuits plays an 
important role in designing QCA logic circuits,\cite{lent1993QCA,lent1993lines,tougaw1993bistable} exploring their robustness against stray charge, layout defects, or other unwanted effects,\cite{cong2022clocked} as well as exploring methods for writing classical bits to--or reading them from--molecular QCA circuits.\cite{blair2019electric-field-input,cong2024circuits}

An exact classical approach to modeling the ground state is to diagonalize a matrix representing the circuit Hamiltonian. While this approach includes coherences between devices, this is challenging because the size of \(\hat{H}\) scales exponentially with the number of cells in the circuit: for a \(d\)-state description of a single QCA cell, the circuit Hamiltonian for an \(N\)-cell circuit is a \(d^{2N}\) matrix, and exact diagonalization is unfeasible for circuits with more than a few cells. 

To model the behaviors of larger circuits other approximations have been applied. The inter-cellular Hartree approximation (ICHA) \cite{lent1993QCA,lent1993lines,tougaw1993bistable} calculates the ground state for individual cells. The influence of the neighbors on the \(k\)-th cell is included as a mean-field, classical contribution to the \(k\)-th cell's occupation energies. The algorithm starts with an initial guess and iterates to convergence, finding a local energy minimum. This method greatly reduces the computational cost of the ground state from diagonalizing a \(d^{2N}\)-element matrix to an repeated diagonalization of \(N\) \(2\times2\) matrices iterated to self-consistency; however, this iterative algorithm is sensitive to the initial guess, and coherence between cells is discarded. The ICHA model yields incorrect results when correlations are significant within a circuit.\cite{toth2001role} Partially-coherent models use coherence vector formalism, where multi-cell correlations may be included as desired, subject to the availability of computational resources. Such models were used to estimate not just the ground state, but also isolated circuit dynamics \cite{toth2001role} and dissipative circuit 
dynamics.\cite{timler2002power,timler2003maxwell,taucer2015consequences} Additionally, quantum annealing \cite{lloyd1996universal} has been explored to model the ground state of QCA 
circuits.\cite{retallick2014embedding} Another method was developed to divide a large QCA circuit into subcircuits and to estimate the overall circuit ground state--as well as other low-energy eigenstates--from approximations of the subcircuit Hamiltonian matrices.\cite{retallick2021low}

In this work, we explore the use of variational quantum eigensolver (VQE) \cite{peruzzo2014variational} methods to the estimation of the ground state of a QCA circuit.


\section{Model} \label{sec:Model}

\subsection{Hamiltonian Formulation}

In the two-state approximation,\cite{tougaw1993bistable} each molecular QCA cell can be effectively reduced from its 
full multi-electron description to a simplified two-level quantum system, represented by the cell's 
binary polarization states. 
Under this approximation, the molecular QCA Hamiltonian takes the form of an Ising-type spin Hamiltonian,\cite{toth2001quantum} directly aligning with standard qubit representations employed in quantum computing. Consequently, QCA circuits under this simplified treatment can be naturally represented on quantum hardware without any additional fermion-to-qubit mappings.\cite{jordan1928paulische,bravyi2002fermionic,seeley2012bravyi} This direct spin (qubit) representation greatly simplifies the encoding of molecular QCA problems into quantum circuits, thus streamlining their implementation and simulation using VQE. 

The Hamiltonian, \(\hat{H}\), for a single QCA cell may be written in the computational basis as
\begin{equation}
\hat{H} = -\gamma \mathbf{X} - \frac{\Delta}{2} \mathbf{Z} ,
\label{eq:single-cell-H}
\end{equation}
where \(\gamma\) is the tunneling energy between the states \(\ket{0}\) and \(\ket{1}\), \(\mathbf{X}\), \(\Delta = \braket{1| \hat{H} | 1 } - \braket{0| \hat{H} | 0}  \) is the bias between states 1 and 0, and \(\mathbf{X} = \sigma_x\) and \(\mathbf{Z}=\sigma_z\) are Pauli operators. 

The Hamiltonian for an \(N\)-cell QCA circuit may be written as follows:
\begin{equation}
\hat{H} = \sum_{n=0}^{N} \hat{H}_n + \sum_{m, n>m} E_{m,n} \mathbf{Z}_{m} \mathbf{Z}_n,
\label{eq:circuit-hamiltonian}
\end{equation}
where \(\hat{H}_n\) is the Hamiltonian of the \(n\)-th cell, and \(\mathbf{Z}_{k}\) is the Pauli-\(z\) operator acting on 
the Hilbert space of the \(k\)-th cell. The first summation is the free Hamiltonian of all cells in the circuit, and the 
double summation includes all pair-wise intercellular interactions. \(E_{m,n}\) is an energy describing the coupling 
between cells \(m\) and \(n\), and it depends on the relative positions of cells \(m\) and \(n\). \(E_{m,n}\) may be 
calculated in a straight-forward way by treating electrons as point charges and summing electrostatic potential energies. We restrict this summation to have \(m>n\) in order to avoid double-counting intercellular interactions. Additionally, for simplicity, we will ignore energies beyond next-nearest neighbors.
In this VQE model of a QCA circuit, each QCA cell corresponds directly to a single qubit in the quantum circuit used to model the QCA network.

\begin{figure*}[htbp]
    \centering
    \subfloat[\label{fig:interaction-horizontal}]{%
      \frame{\includegraphics[width=0.235\textwidth]{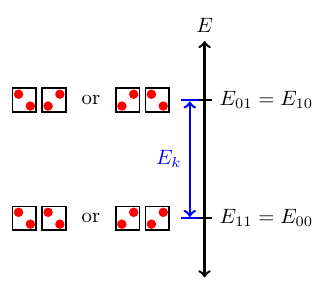}}%
    } \quad
    \subfloat[\label{fig:interaction-diagonal-up}]{%
      \frame{\includegraphics[width=0.235\textwidth]{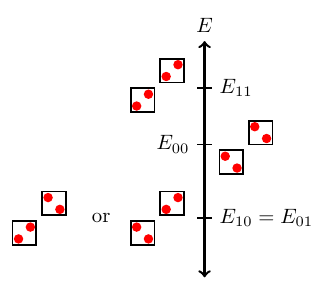}}%
    } \quad
    \subfloat[\label{fig:interaction-diagonal-down}]{%
      \frame{\includegraphics[width=0.235\textwidth]{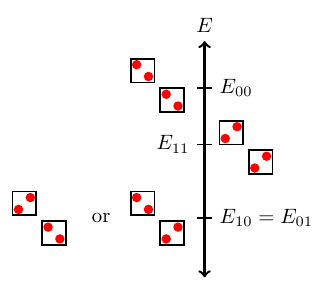}}%
    } \quad
    \subfloat[\label{fig:interaction-diagonal-simple}]{%
      \frame{\includegraphics[width=0.235\textwidth]{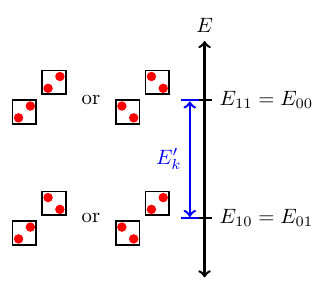}}%
    } \quad
    \caption{While nearest-neighbor interactions are straight-forward, we use a simplified description of next-nearest-neighbor interactions. (a) Aligned states are favored for wo cells coupling horizontally as nearest-neighbors. Aligned interactions are degenerate and lie \(E_k\) below anti-aligned interaction energies, which also are degenerate. These same relationships hold for vertically-arranged nearest-neighbors (not shown here).
      (b) Diagonal (next-nearest-neighbor coupling) favors anti-aligned states, ``10'' or ``01,'' over the aligned states ``00'' or ``11.'' In a ``forward-slash'' configuration, the ``11'' state has a higher interaction energy than does the ``00'' state. (c) In a ``back-slash'' configuration, diagonal coupling again favors the anti-aligned states, but the "00" state is more energetic than the ``11'' state. (d) In this work, we use a simplified description of next-nearest-neighbor interactions, in which both the anti-aligned interactions and the aligned interactions are degenerate, with a new diagonal kink energy, $E_{k}^{\prime}$.}
    \label{fig:1-cell-layout}
\end{figure*}

Nearest-neighbor interactions may be described in a straight-forward manner. Horizontal interactions in the standard basis are shown in Figure \ref{fig:interaction-horizontal}, where \(E_{\beta \alpha}\) is the interaction energy between cell 0 (the left cell) in logical state $\alpha$ and cell 1 (on the right) in logical state $\beta$. Generally, we calculate any interaction energy by setting the cells in their respective states, treating the charges within each cell as point charges, all summing over all pairwise intercellular Coulomb energies. The interaction energy is minimized at $E_{00} = E_{11}$ when the two cells are aligned, and maximized at $E_{10} = E_{01}$ when they are anti-aligned. This maps directly onto vertical interactions, which are not shown here.
The anti-aligned states are degenerate and are called ``kinked'' states in QCA. The energy difference between the kinked states and the aligned states is called the kink energy, \(E_k = E_{10} - E_{11}\), and this is interpreted as the cost of a bit flip. The degeneracy of the kinked states, along with the degeneracy of the relaxed (aligned) two-cell states, makes it straightforward to write the interaction between the nearest-neighbor pair as a product of \(\mathbf{Z}\) operators:
\(\hat{H}_{b,a} = E_k \mathbf{Z}_b \otimes \mathbf{Z}_a\). Thus, $E_{m,n} = E_k$ in Equation (\ref{eq:circuit-hamiltonian}) for horizontal or vertical nearest-neighbor interactions. In this work, we assume $a= 1$ nm, which results in $E_k = -294.3$ meV. The fact that $E_k < 0$ tends to cause cells to align through nearest-neighbor interactions.

The next-nearest-neighbor interactions are diagonal, as illustrated in Figures \ref{fig:interaction-diagonal-up} and \ref{fig:interaction-diagonal-down}. These interactions are more complicated than the horizontal or vertical interactions. In this case, the anti-aligned states have the lowest interaction energies, and they are degenerate as before: \(E_{10}=E_{01}\). The excited states are the aligned states. In the case of Figure \ref{fig:interaction-diagonal-up}, $E_{11} > E_{00}$; but, in the case of Figure \ref{fig:interaction-diagonal-down}, $E_{00} > E_{11}$. In this work, we use a simplifying treatment: we neglect the energy difference between the aligned states, so that $E_{11}=E_{00}$ as in Figure \ref{fig:interaction-diagonal-simple}. This allows us to write the interaction once again as a product of $\mathbf{Z}$ operators, this time using another energy, $E^{\prime}_k = E_{00} - E_{01}$: \(\hat{H}_{b,a} = E^{\prime}_k \mathbf{Z}_b \otimes \mathbf{Z}_a\). Thus, $E_{m,n} = E^{\prime}_k$ in Equation (\ref{eq:circuit-hamiltonian}) for diagonal nearest-neighbor interactions. With $a= 1$ nm, we obtain $E_{11} - E_{01} =  85.7$ meV and $E_{00} - E_{10} =  45.6$ meV for the configuration of Figure \ref{fig:interaction-diagonal-up}. The positive value $E^{\prime}_k >0$ favors anti-alignment in next-nearest-neighbor pairs. In this work, we choose for simplicity $E^{\prime}_{k} = 85.7$ meV.

\subsection{Variational Quantum Eigensolver}

\begin{figure*}[htbp]
  \centering
  \includegraphics[width=0.7\textwidth]{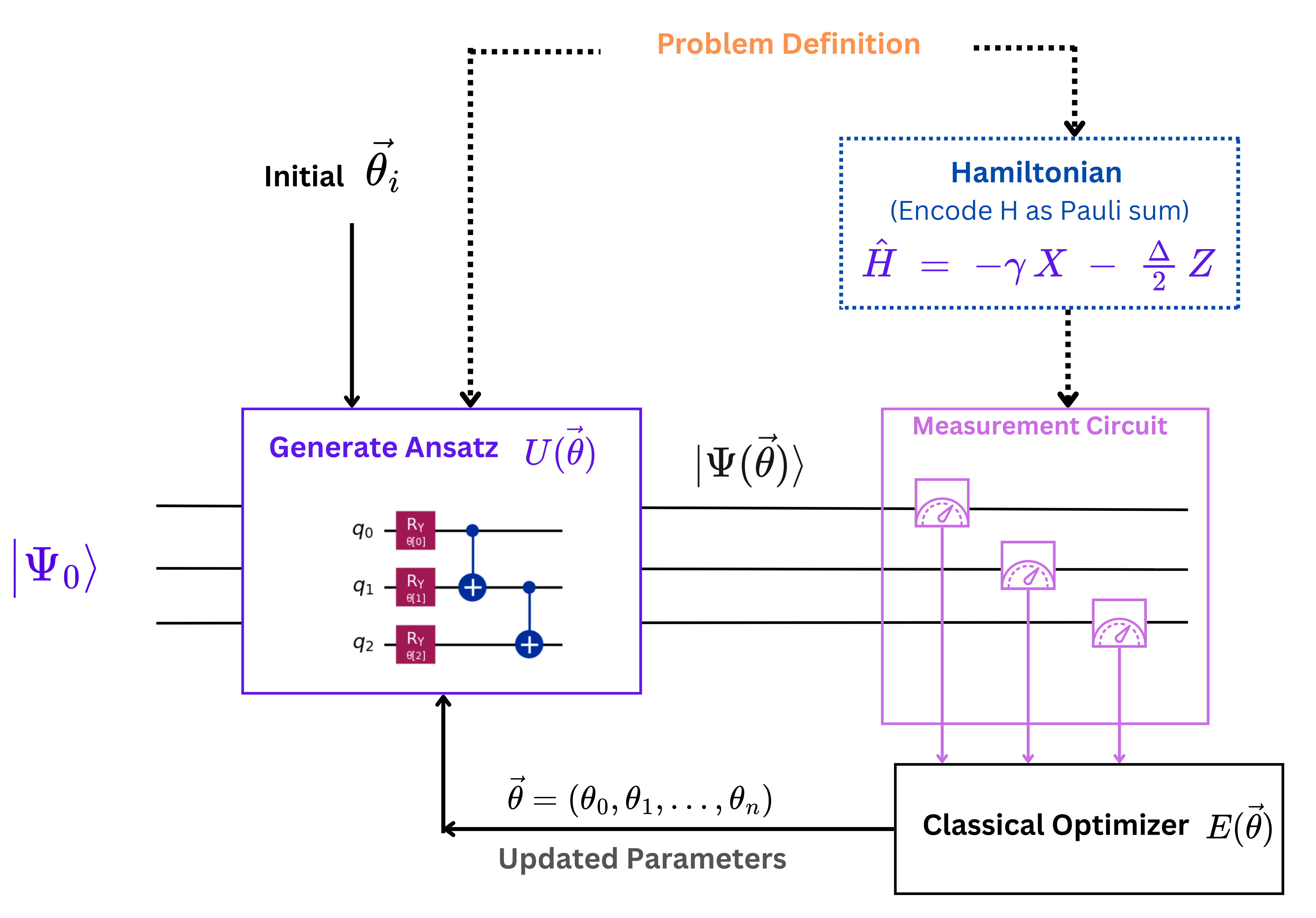}
  \caption{The VQE workflow begins with the construction of a parameterized 
  quantum circuit (ansatz), which prepares a trial quantum state. 
  This state is used to estimate the expectation value of the 
  Hamiltonian via quantum measurements. The resulting energy estimate 
  is then fed into a classical optimizer, which updates 
  the circuit parameters to iteratively minimize the energy. This hybrid quantum-classical 
  feedback loop continues until convergence to the approximate ground-state energy is achieved.}
  \label{fig:VQEblock}
\end{figure*}

The Variational Quantum Eigensolver (VQE) \cite{peruzzo2014variational} algorithm estimates the ground-state energy \( E_0 \) 
of a quantum system by variationally optimizing a parameterized trial 
wavefunction \( \ket{\Psi(\vec{\theta})} \) and estimating its state energy. The overall process is shown in Figure~\ref{fig:VQEblock}.

The objective of VQE is to minimize the expectation value of the system Hamiltonian with respect to the trial state:
\begin{equation}
E(\theta) = \Braket{ \Psi (\vec{\theta} ) | \hat{H} | \Psi (\vec{\theta}) },
\end{equation}
where \( \ket{ \Psi(\vec{\theta})} = U(\vec{\theta}) \ket{ \Psi_0 } \) is the trial wavefunction. 
The Hamiltonian determines a set of measurements required to estimate the energy of \( \ket{ \psi(\theta) } \). Additionally, we must choose a parametric ansatz circuit, \(\mathbf{U} (\vec{\theta} )\), which can prepare wavefunctions \(\ket{\Psi}\), which correspond to a quantum state in the neighborhood of the ground state from a starting state, \(\Ket{\Psi_0}\). Typically, the starting state is prepared as \(\Ket{\Psi_0} = \ket{0}\). Given expectation values from measurement, the classical optimizer routine estimates \(E(\vec{\theta})\) and employs an algorithm to minimize \(E(\vec{\theta})\) with respect to the circuit parameters, \(\vec{\theta} = (\theta_0, \theta_1, \theta_2, \ldots) \). The optimizer adjusts the parameter vector \(\vec{\theta}\), modifying the ansatz to create a new trial function and a new estimate of the energy of the quantum system under study. The process repeats until a threshold of convergences. Formally, the objective is to find the optimal parameters \( \vec{\theta}^* \) that minimize the energy expectation value:
\begin{equation}
\vec{\theta}^* = \arg\min_{\vec{\theta}} \Braket{ \Psi(\vec{\theta}) | \hat{H} | \Psi(\vec{\theta}) },
\end{equation}
yielding an estimate of the ground-state energy:
\begin{equation}
E_0 \approx E(\vec{\theta}^*) = \Braket{ \Psi(\vec{\theta}^*) | \hat{H} | \Psi(\vec{\theta}^*) }.
\end{equation}

\subsection{Circuit Response}

The QCA circuit response is modeled in a straight-forward way after the ground state is estimated and  \(\vec{\theta}^{\ast}\) is obtained. The polarization of any QCA cells may simply be found by applying \(U(\vec{\theta}^{\ast})\) once again to \(\Psi_0\) and finding the expectation value of a measurement in the \(\mathbf{Z}\) basis on the qubits corresponding to the QCA cells of interest.

\subsection{Computational Resources}

Whenever possible, exact classical results were obtained as a benchmark for VQE results. VQE was simulated using Qiskit's \cite{qiskit2024} AerSimulator, or it was performed 
on cloud-based hardware available through IBM.\cite{IBM2025} For this work, the quantum processors used are listed in Table \ref{tab:qpus}. The choice of hardware used for calculations was determined in part by availability of resources. \verb|ibm_sherbrooke| was used for the one-cell and three-cell binary wires, and the two-cell majority gate. All other circuits were modeled using the newer \verb|ibm_kingston|. \verb|ibm_kingston| affords more qubits with an improved noise profile, lending itself to more accurate VQE models of QCA than does \verb|ibm_kingston|.

\begin{table}[htp]
\caption{The following IBM quantum processing units (QPUs) were used for this work.}
\begin{center}
\begin{tabular}{|c|c|c|c|}
\hline
\textbf{QPU} & \textbf{Model} & \textbf{Logical qubits} & \textbf{Lowest 2Q error} \\
\hline
\verb|ibm_sherbrooke| & Eagle r3 & 127 & $2.33 \times 10^{-3} $\\
\hline
\verb|ibm_torino| & Heron r1 & 133 & $1.57 \times 10^{-3}$ \\
\hline
\verb|ibm_kingston| & Heron r2 & 156 & $8.67 \times 10^{-4}$ \\
\hline
\end{tabular}
\end{center}
\label{tab:qpus}
\end{table}%

\section{Results} \label{sec:Results}

\subsection{Binary Wire}

\subsubsection{Single-Cell Binary Wire}

\begin{figure}[htbp]
  \centering
  \begin{minipage}[t]{0.475\columnwidth}
    \centering

    \subfloat[\label{subfig:BinaryWire1}]{%
      \includegraphics[width=0.8\textwidth]{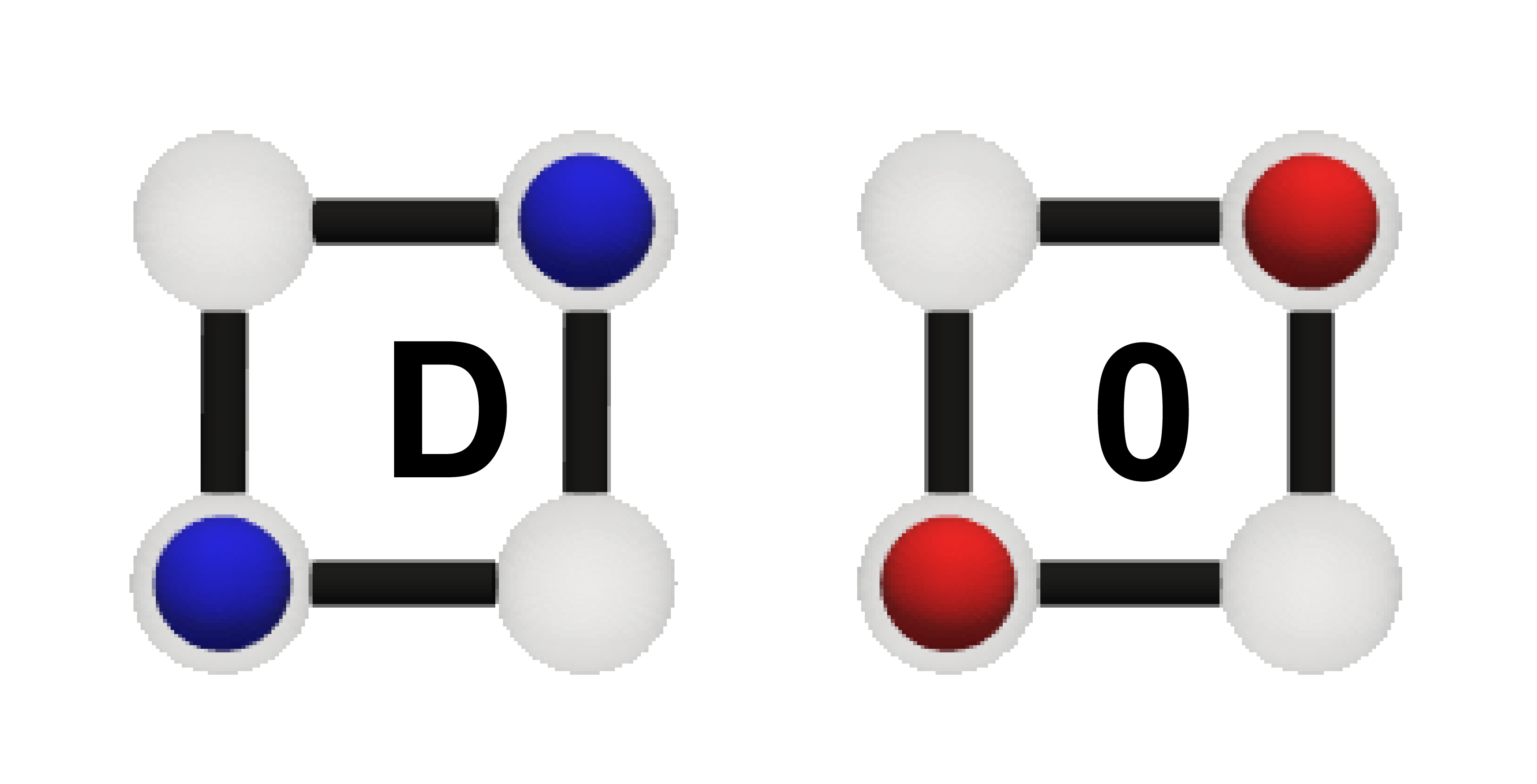}%
    } \\
    \subfloat[\label{subfig:1-ansatz}]{%
      \includegraphics[width=0.6\textwidth,trim=1cm 0.5cm 0 1cm, clip]{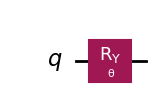}%
    }
  \end{minipage}
  \hfill
  \begin{minipage}[t]{0.475\columnwidth}
    \centering
    \subfloat[\label{fig:bloch-ryonly}]{%
      \includegraphics[width=\linewidth,trim=18.5cm 2.5cm 15cm 5cm, clip]{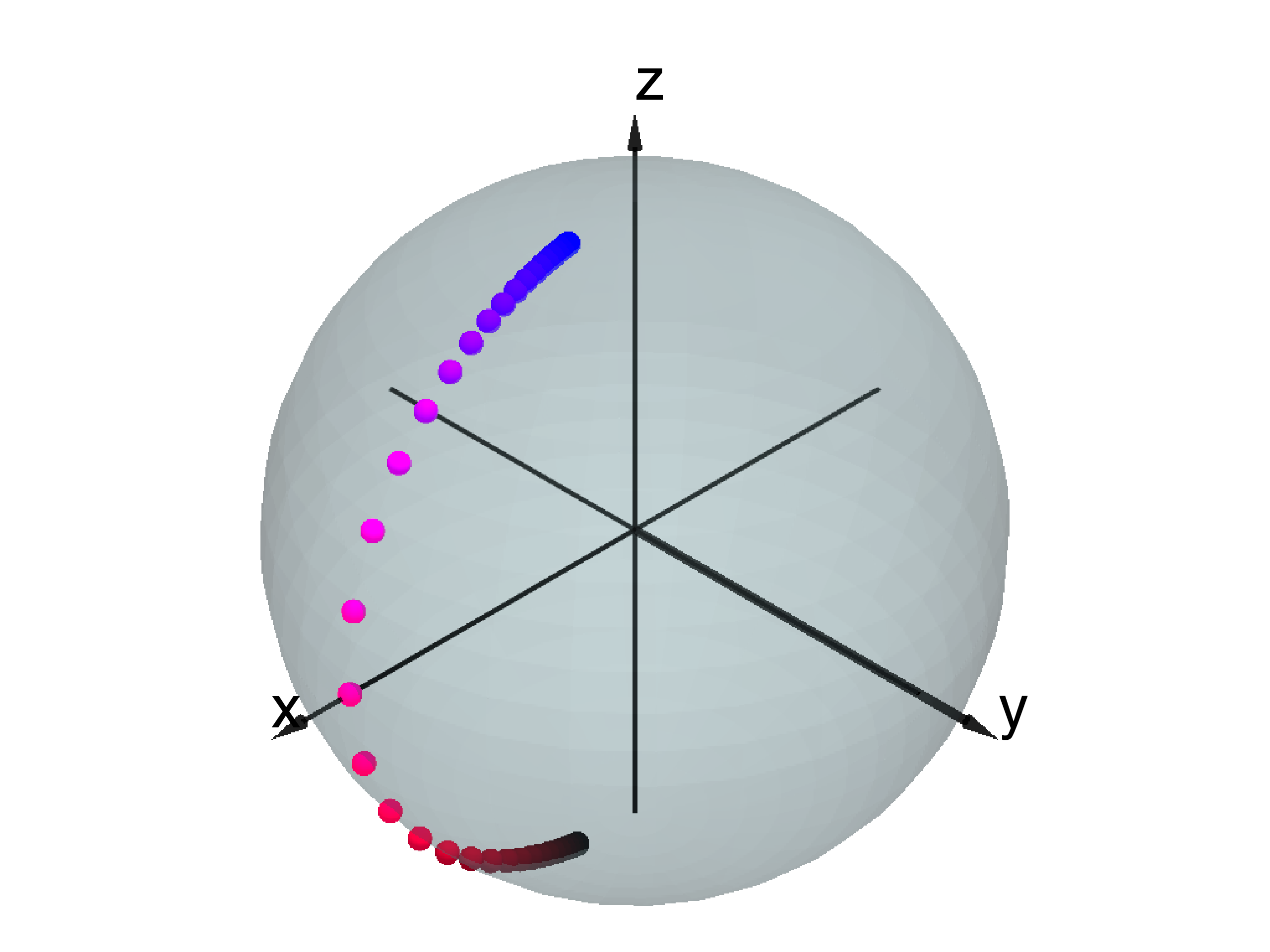}
      \label{fig:right}
    }
  \end{minipage}

  \caption{A single \(\mathbf{R}_y(\theta)\) gate may serve as an ansatz for a driven four-dot QCA cell. (a) A single QCA cell (right) tends to align with a driver by when coupled via a nearest-neighbor interaction. 
(b) Ansatz used for the VQE simulation of a 1 cell binary wire shows
the parameterized quantum circuit that is used to prepare the initial state of the
QCA cells. The circuit only has a single parameter (single theta value), which minimizes the complexity of the classical optimization.
(c) The ground state, \(\ket{\psi (P_{drv}) }\) for the driven cell QCA cell lies within the \(xz\) plane in the Bloch sphere over the range of valid \(P_{drv}\) values. Thus, a \(\mathbf{R}_y(\theta)\) gate is sufficient to transform \(\ket{0}\) into \(\ket{\psi (P_{drv})}\). These vectors shown here are calculated from the exact
diagonalization of \(\hat{H}\) over the range of valid driver polarizations, \(P_{drv}\).}
  \label{fig:single-cell-model}
\end{figure}

We begin by modeling the shortest binary wire possible: a single QCA cell influenced by a driver cell, as depicted in Figure \ref{subfig:BinaryWire1}.
The ansatz selected is shown in  Figure \ref{subfig:1-ansatz}. Only a single qubit is required to model the quantum state of the single, driven QCA cell, and a single y-rotation gate, \(\mathbf{R}_y\) is used to achieve the trial wavefunction. This was chosen primarily because it is sufficient to transform the initial state \(\ket{0}\) into the ground state wave, and second because of its simplicity, having with only a single parameter for ease of optimization.

The sufficiency of the ansatz for this system may be demonstrated by diagonalizing the Hamiltonian of Equation (\ref{eq:single-cell-H}). We may write the bias \(\Delta\) in Equation (\ref{eq:single-cell-H}) as a function of physical constants, the driver polarization, \(P_{drv}\), and the cell length parameter, \(a\):
\begin{equation}
  \Delta = \frac{q^2 P_{\text{drv}}}{4\pi \varepsilon_0 a}
\left( -\frac{1}{3} - \frac{2\sqrt{2} - \sqrt{5} - 1}{\sqrt{10}} \right),
\end{equation}
With \(a = 1\) nm and \(\gamma = 50\) meV, we diagonalize \(\hat{H}\) to find the ground state, calculate the Bloch vector, \(\vec{\lambda}\), and then plot the tip of \(\vec{\lambda}\) as a function of driver polarization over its range of valid values, \(P_{drv} \in \left[ -1, 1\right] \). The results shown in Figure \ref{fig:bloch-ryonly} all lie on an arc within the \(xz\) plane of the Bloch sphere. Thus, these points are accessible using only rotations  of \(\ket{0}\) (pointing in the \(+z\) direction) about the \(y\) axis in the Bloch sphere. These rotations may be implemented using a parametrized \(\mathbf{R}_y(\theta)\) gate, with angle \(\theta\) specifying the angle of rotation.

We found the Constrained Optimization by Linear Approximation (COBYLA) \cite{pellow2021comparison}
optimizer from Python's SciPy library to provide the most rapid and flexible classical optimization for our circuits. 

\begin{figure}[htbp]
  \centering
  \includegraphics[width=0.5\textwidth]{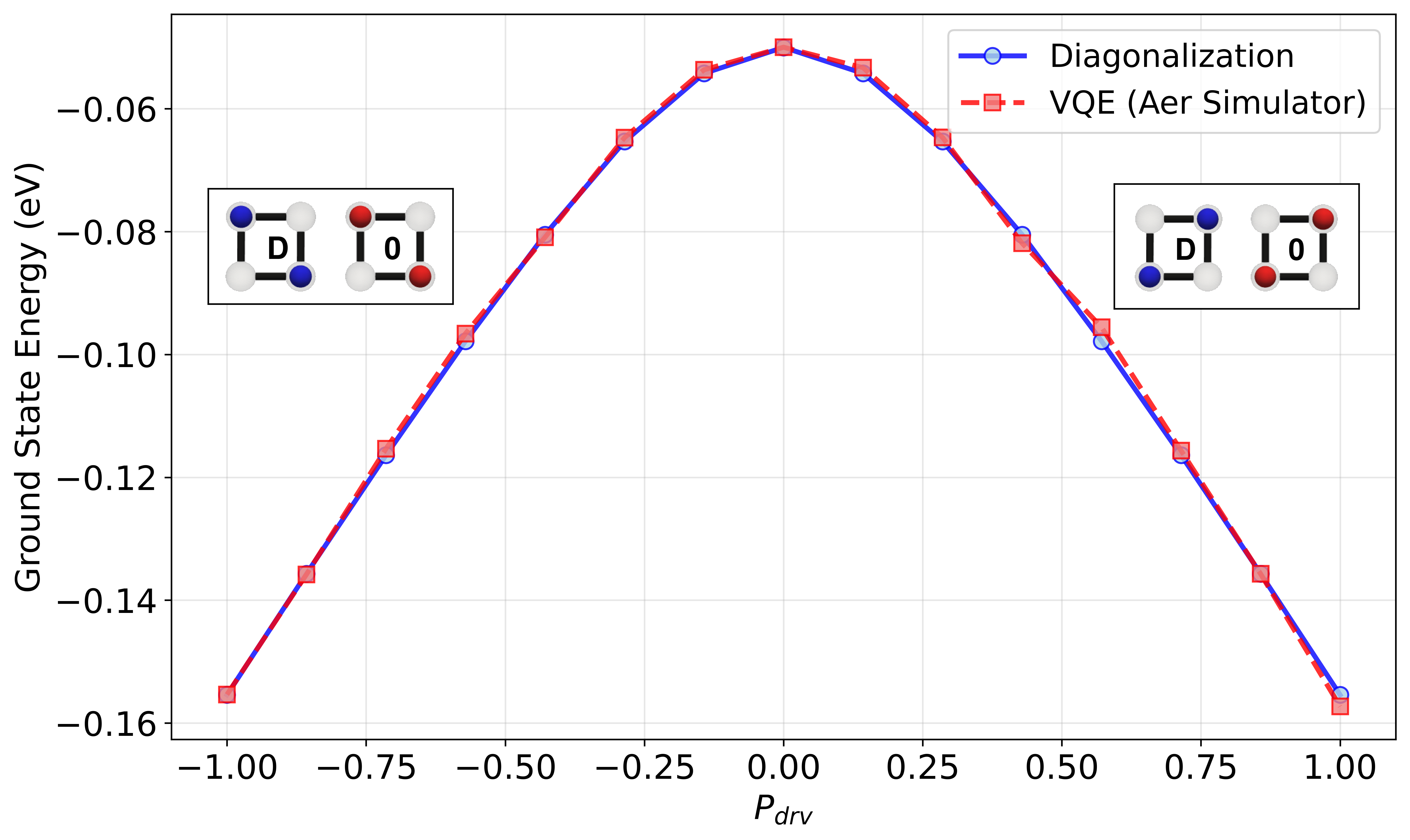}
  \caption{A noise-free VQE simulation estimates the ground state of a single-cell QCA binary wire as a function of driver polarization, \(P_{drv}\). Noise-free VQE simulation results match exact results closely, within \(k_BT\) at room temperature.
  }
  \label{fig:1-cell-energy}
\end{figure}

Figure \ref{fig:1-cell-energy} shows data for a noise-free, simulated VQE estimation of the driven single-cell binary wire. The energy is shown as a function of the driver molecule's polarization, \(P_{drv}\).  The noise-free VQE simulation agrees well with the ground state energy from the exactly-diagonalized single-cell wire Hamiltonian to within fractions of \(k_B T\) at room temperature. While this provides excellent validation of the single-cell VQE model in the limit of noise-free simulation, this does not readily demonstrate proper device operation. To better illustrate that the molecular circuit's ground state from VQE exhibits the desired behavior, we prefer to show the polarization response of the QCA cells. This means taking the process one step beyond VQE: after the ground state is approximated through VQE, each qubit of the optimized circuit is again sampled in the computational basis to estimate cell polarizations.

\begin{figure}[htbp]
  \centering
  \includegraphics[width=0.5\textwidth]{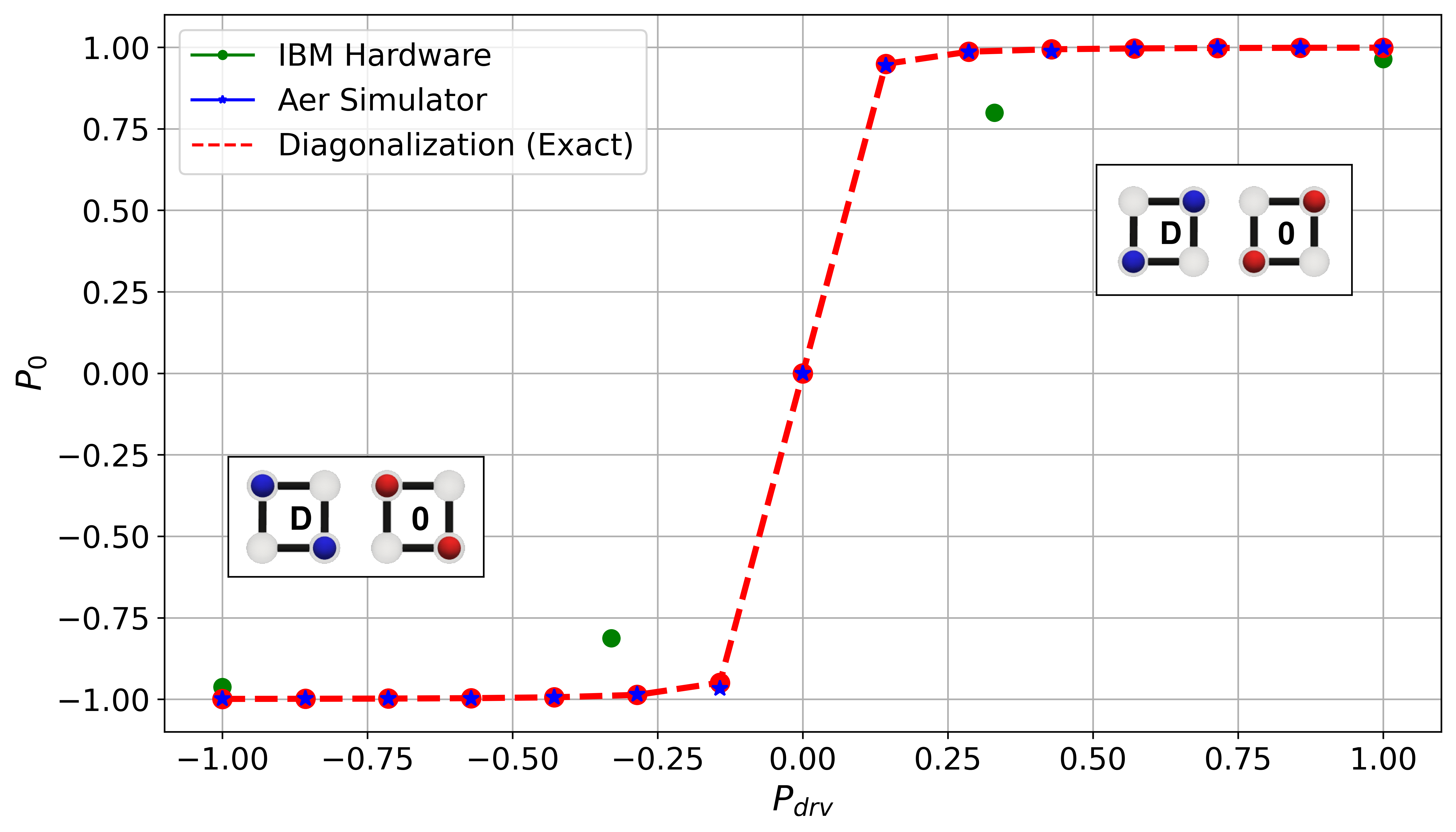}
  \caption{Measurements on the VQE result accurately approximate the cell-cell response
  for a single QCA cell as the function of driving 
  polarization. Exact solutions using 
  matrix calculations and VQE results for both noiseless (AER) and noisy (IBM Hardware)
  show agreement,  however with somewhat reduced accuracy for the noisy one. 
  This response is seen to be non-linear and saturates near 
  the levels P = \(\pm\)1 
  showing that the system is bistable.  
  }
  \label{fig:1-qca-all}
\end{figure}

Model results for cell polarization are shown for the single-cell wire in Figure \ref{fig:1-qca-all}. Here, noiseless Aer Simulator results match results from the exact diagonalization of \(\hat{H}\) very closely, providing validation of the chosen ansatz and the VQE circuit description. The trend shown here is the cell-cell polarization curve that is desirable in QCA. The response is non-linear and exhibits a signal gain, so that even a weak polarization in the driver cell leads to a strong polarization in the target cell. The hardware results capture the same response as the exact calculation and simulated VQE results, however with somewhat reduced accuracy. In this and following sets of hardware results, we include much fewer data points due to the scarcity of resources and the need to conserve hardware runtime. Additionally, to conserve hardware runtime, we used only the default precision setting, which directly controls the number of shots in a given experiment. A higher precision and more shots would likely improved the accuracy of the results. Overall, these results indicate the suitability and effectiveness of our approach for modeling quantum-dot cellular automata using VQE methods.

\begin{figure}[htbp]
  \centering
  \includegraphics[width=0.35\textwidth]{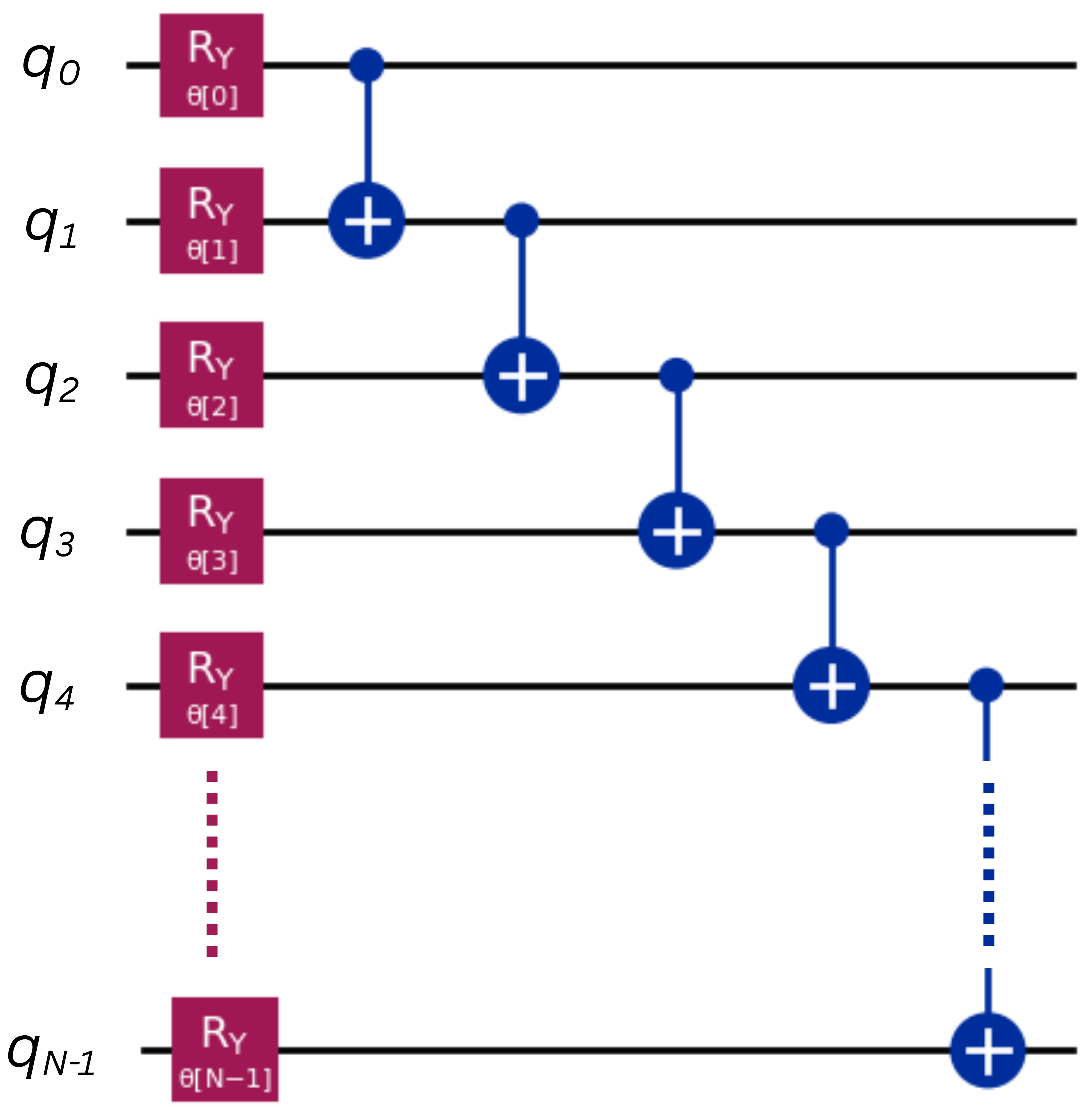}
  \caption{A \(N\)-qubit extension of the ansatz of Figure \ref{subfig:1-ansatz} may be used to model an \(N\)-cell binary wire.
  Here, entangling gates \(\mathbf{C}_{k,k+1}\) couple adjacent cells, and single-qubit rotations \(\mathbf{R}_y(\theta_k)\) allow arbitrary
  polarizations on each cell. This structure suffers from increased optimization complexity as the number of 
  parameters grows with \(N\). For practical scalability, a reduced \(N\)-qubit ansatz with fewer \(\mathbf{R}_y\) gates may be used. }
  \label{fig:N-cell-full-ansatz}
\end{figure}

\begin{figure}[htbp]
  \centering
  \includegraphics[width=0.45\textwidth]{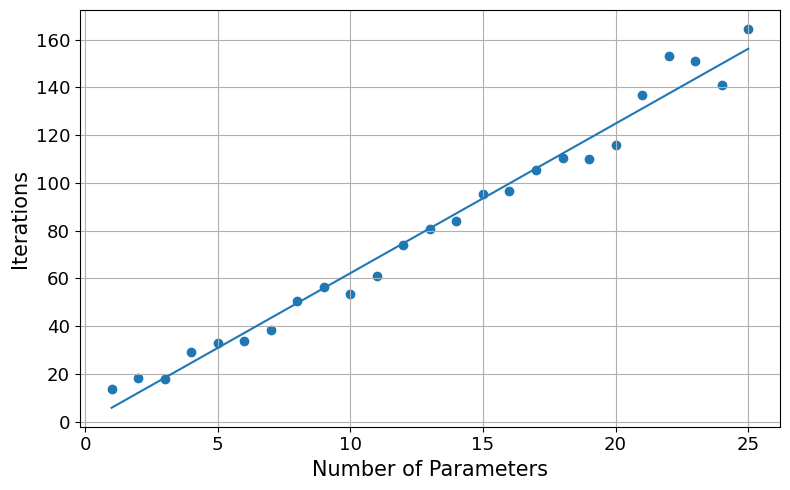}
  \caption{The number of COBYLA optimizer iterations required in a simulated VQE model grows linearly with the number of of parameters in the parametric circuit. The dots show actual results, and the solid line is a linear fit. In this case, the number of parameters corresponds to the number of cells in a binary wire, with a single \(\mathbf{R}_y(\theta)\) gate associated with each qubit, as in Figure \ref{fig:N-cell-full-ansatz}.}
  \label{fig:paramsviter}
\end{figure}

Ideally, to extend the binary wire to \(N\) cells, we would add a controlled-NOT gate, \(\mathbf{C}_{k,k+1}\), with the \(k\)-th qubit as the control qubit, and the \((k+1)\)-th qubit as the target bit. This couples the bit of the \(k\)-th QCA cell to the \((k+1)\)-th cell. Additionally, we would add an \(\mathbf{R}_y(\theta_k)\) gate to the \(k\)-th qubit, as shown in Figure \ref{fig:N-cell-full-ansatz}. This would allow for arbitrary bit flips (kinks) along the wire. One challenge here is that the complexity of the 
COBYLA optimzation grows with the number of parameters in the classical optimization, as seen in Figure \ref{fig:paramsviter}. Here,
the number of VQE iterations required to obtain a given data point grows linearly with the number of parameters, which--in this case--corresponds to the
number of QCA cells in the binary wire. While the plot of Figure \ref{fig:paramsviter} was obtained using noise-free simulations,
we can expect this growth (or perhaps worse) on noisy hardware, and this will correlate directly to increased runtime on the
quantum hardware.
Thus, we prefer models of QCA circuits that minimize the number of optimization parameters. We will refer to an ansatz that eliminates some of the \(\mathbf{R}_y(\theta)\) gates as a reduced ansatz.

\subsubsection{3-Cell Binary Wire}

\begin{figure}[htbp]
    \centering
    \subfloat[\label{fig:1111}]{%
      \includegraphics[width=0.2\textwidth]{img/1111.png}%
    }
    \subfloat[\label{fig:3-ansatz}]{%
      \includegraphics[width=0.2\textwidth]{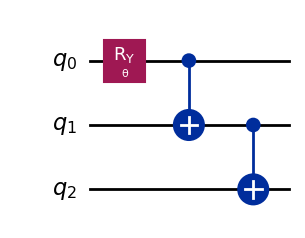}%
    }
    \caption{A single cell binary wire can be extended to a three-cell binary wire without increasing the parameter complexity as a single parameter is enough to model the circuit. (a) An input bit drives a three-cell binary wire.
      (b) A three-qubit, one-parameter ansatz was chosen for a VQE model of the driven, three-cell binary wire.}
    \label{fig:3-cell-layout}
\end{figure}

\begin{figure*}[htbp]
  \centering
  \includegraphics[width=\textwidth]{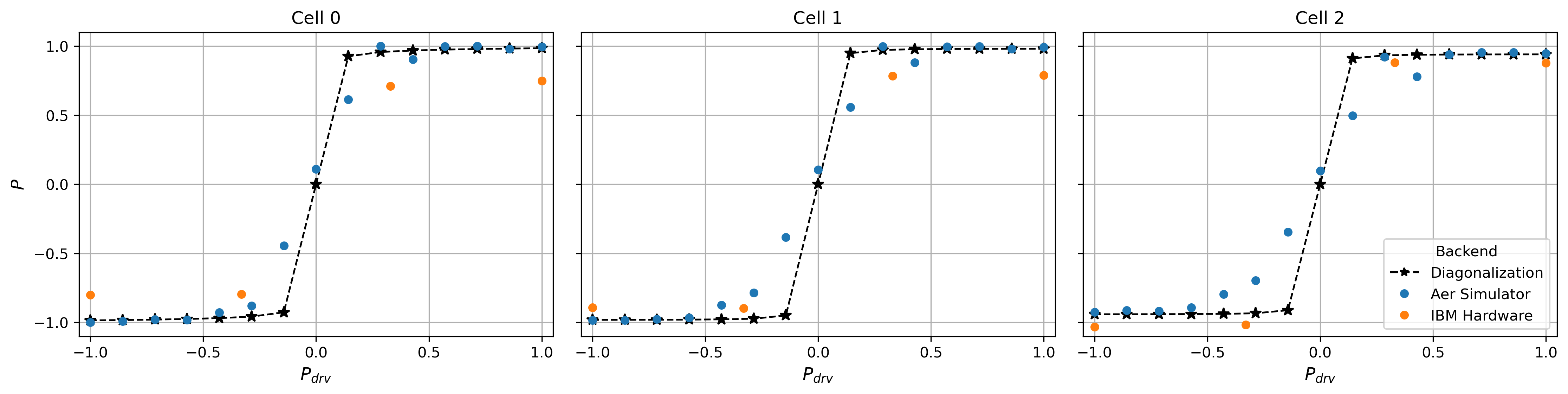}
  \caption{A VQE model captures the transfer of the input bit along a three-cell binary wire, with the AerSimulator closely matching the exact 
    results. While the noisy backend introduces some deviation, it still preserves the 
    overall polarization pattern under realistic quantum noise. }
  \label{fig:3-qca-all}
\end{figure*}

\begin{figure*}[htbp]
    \centering
    \subfloat[\label{fig:3-qca-aer-shots}]{%
      \includegraphics[width=\textwidth]{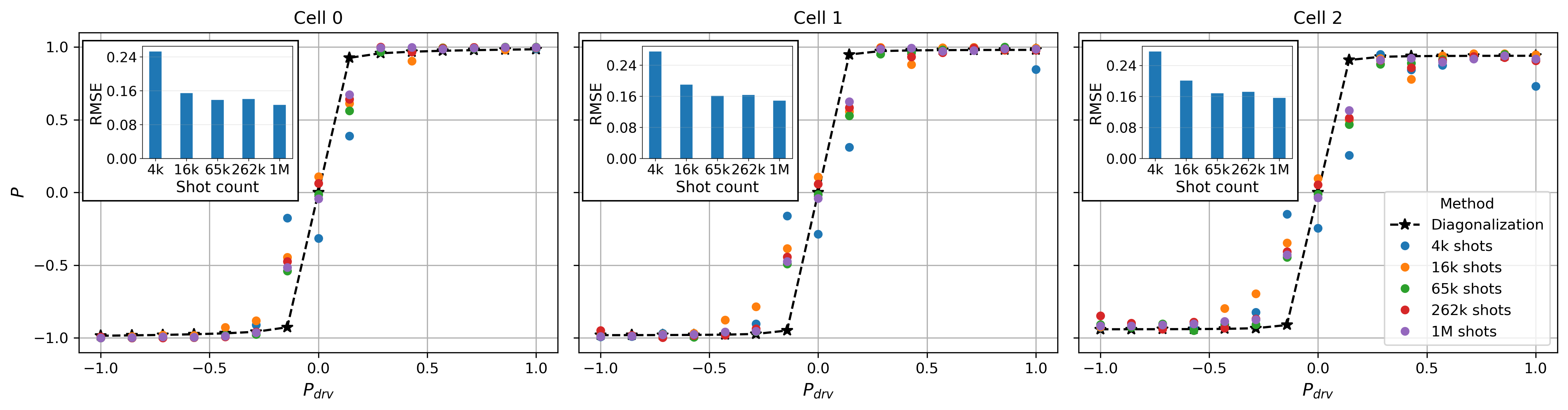}%
    }
\hfill
    \subfloat[\label{fig:3-qca-fake-shots}]{%
      \includegraphics[width=\textwidth]{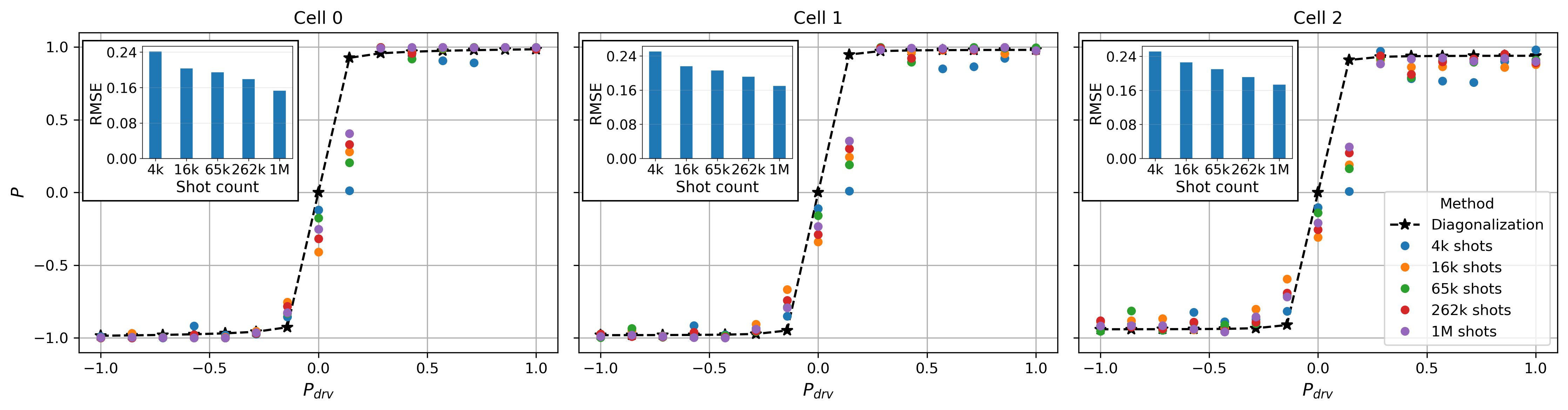}%
    }
\hfill
    \subfloat[\label{fig:3-qca-hardware-shots}]{%
      \includegraphics[width=\textwidth]{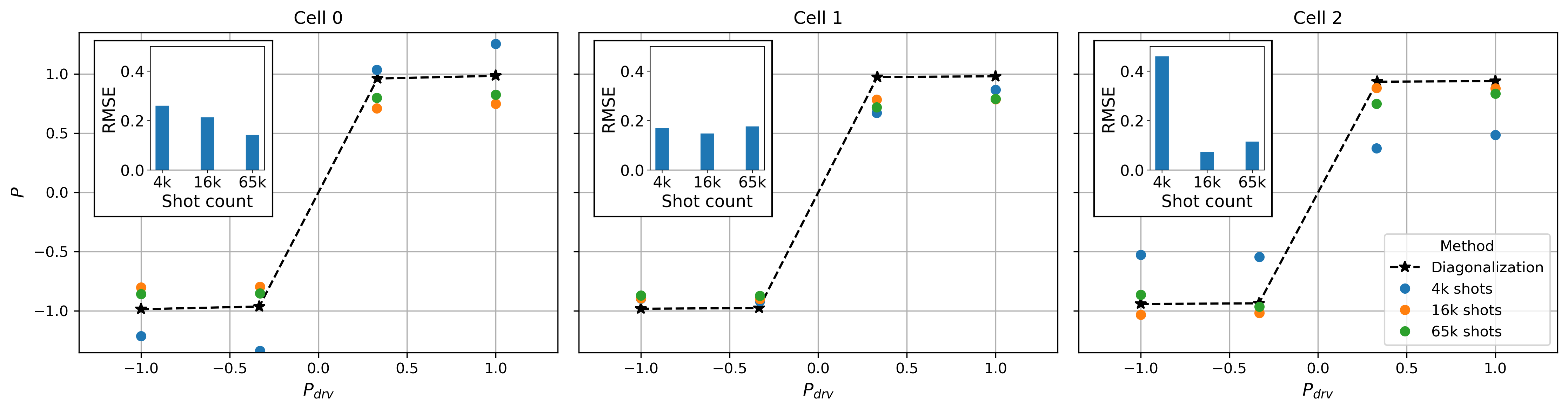}%
    }
    \caption{Increasing the number of shots in energy estimations improves the accuracy of the VQE model. This trend is observed for a three-cell binary wire in noisefree VQE simulation (a), noisy simulation (b) and IBM Hardware (c). To quantify accuracy, a detailed error analysis is presented in Figure \ref{fig:rms-error}.}
    \label{fig:3-cell-shots-comparison}
\end{figure*}

\begin{figure}[htbp]
    \centering
    
    \includegraphics[width=0.48\textwidth]{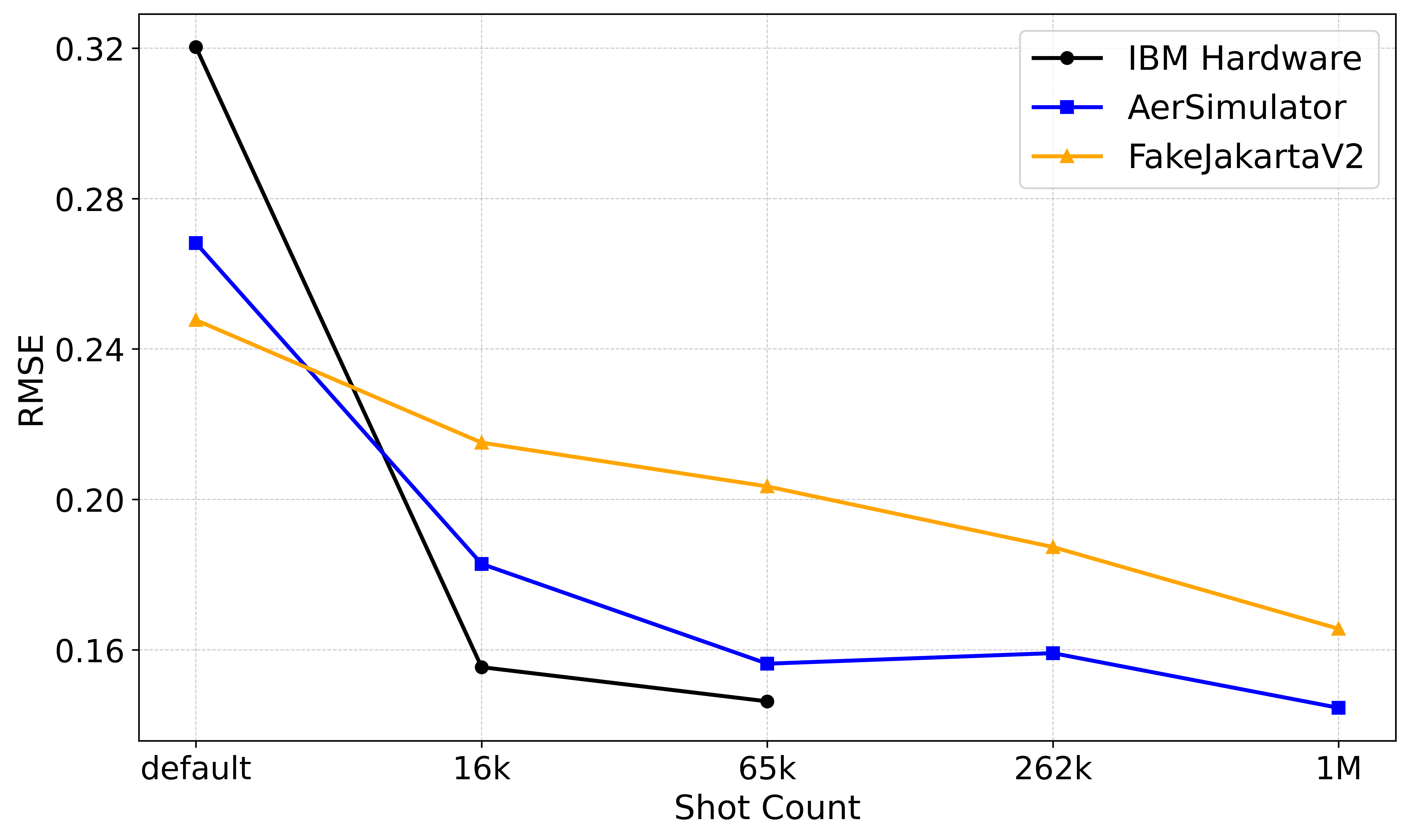}
    
    \caption{16k shots provides a practical trade-off between precision and resource 
      usage for VQE simulations on small QCA circuits.
	The RMSE analysis of the data in Figure \ref{fig:3-cell-shots-comparison}
      demonstrates that increasing the number of shots generally improves accuracy; however, the improvement in RMSE begins to plateau beyond approximately 16k shots, indicating 
      diminishing returns in accuracy despite increased computational cost.}
    \label{fig:rms-error}
\end{figure}

We then extended the binary wire by adding two additional cells, resulting in the three-cell binary wire of Figure \ref{fig:1111}.
The three-qubit ansatz of Figure \ref{fig:3-ansatz} was used. Instead of including a separate \(\mathbf{R}_y(\theta)\) for each qubit,
as in Figure \ref{fig:N-cell-full-ansatz}, we use only a single parameter, \(\theta\), to minimize the complexity of optimization.
The results of this simulation are summarized in Figure \ref{fig:3-qca-all}.
The noiseless Aer Simulator results match the exact diagonalization very closely, demonstrating that the ansatz of Figure \ref{fig:1111} is sufficient to estimate the ground state of the circuit of Figure \ref{fig:3-ansatz}. The hardware results capture the same
trend as the exact and simulated results, however with somewhat reduced accuracy. Notably, the quantum hardware estimated unphysical polarization values for cell 2 outside the valid range \(P_0 \in [-1, 1]\). Additionally, the magnitude of cell polarizations varies along the length of the wire in a manner inconsistent with the exact diagonalization.

These deficiencies are improved in part by performing calculations on newer, less-noisy hardware, or increasing the accuracy setting, which increases the number of shots in each measurement. The increase in accuracy is demonstrated using noise-free simulations in Figure \ref{fig:3-qca-aer-shots} and noisy VQE simulations with FakeBackend in Figure \ref{fig:3-qca-fake-shots} and IBM Hardware in Figure \ref{fig:3-qca-hardware-shots}. In each of these cases, increasing accuracy on the same simulated platform reduces the root-mean-square error (RMSE) of the VQE results relative to the exact results as shown in Figure \ref{fig:rms-error}. 
While increasing the number of shots improves the RMSE, these gains diminish with increasing shot count. Thus, an accuracy corresponding to 16k shots was chosen as a practical balance between time and precision in subsequent VQE simulations. However, only the default number of shots (4096) were used on real IBM hardware to conserve runtime. 



\subsubsection{7-Cell Binary Wire}

\begin{figure}[htbp]
    \centering
    \subfloat[]{%
      \includegraphics[width=0.45\textwidth]{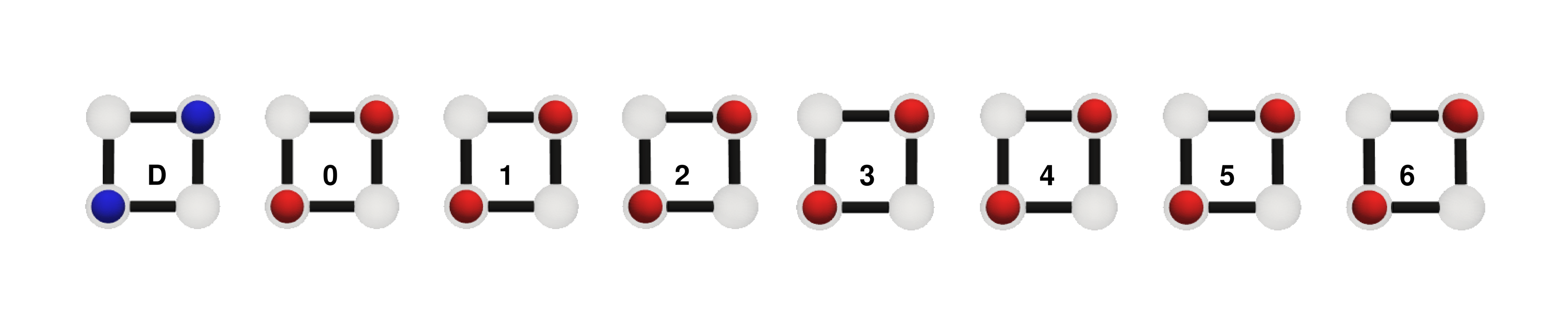}%
      \label{fig:0-6}%
    }
    \vfil
    \subfloat[]{%
      \includegraphics[width=0.3\textwidth]{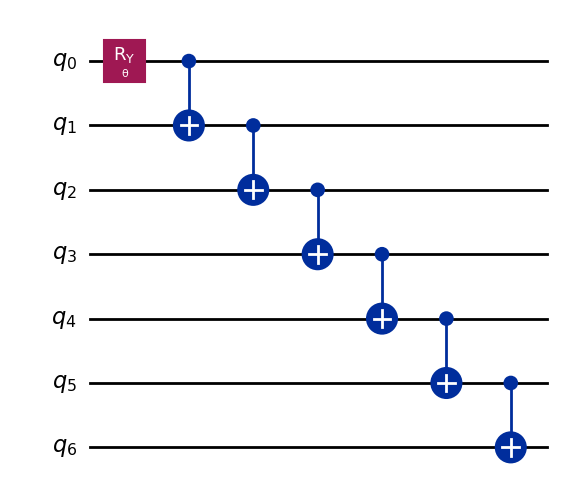}%
      \label{fig:7-ansatz}%
    }
    \caption{A seven-cell binary wire can be modelled with the same number of parameter as a single cell wire. (a) An input bit drives a seven-cell binary wire. 
      (b) The reduced ansatz for a seven-cell wire 
      only has a single parameter and the 
      optimization routine for a 7-cell QCA circuit is similar as the single cell 
      wire or the three-cell one.}
    \label{fig:7-cell-layout}
\end{figure}

\begin{figure*}[htbp]
  \centering
  \includegraphics[width=\textwidth]{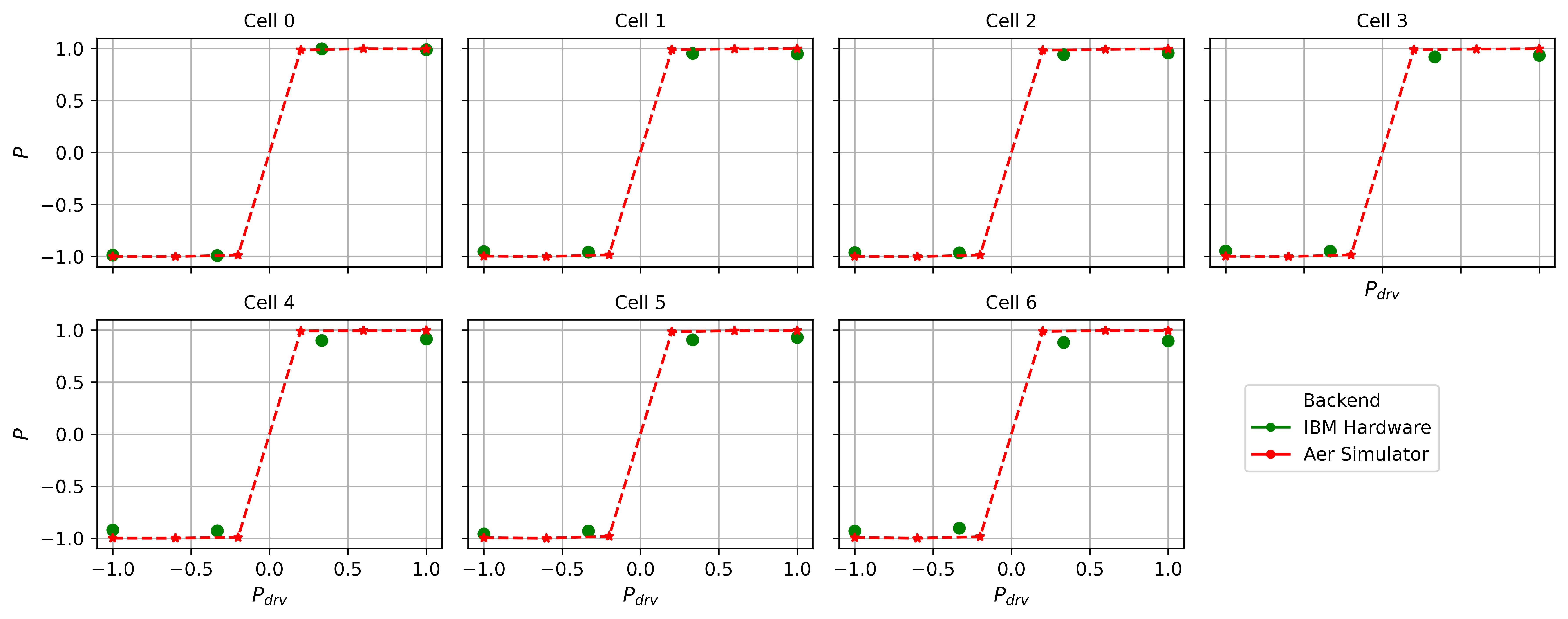}
  \caption{All seven cells demonstrate consistent and aligned 
    polarization behavior, indicating successful information propagation along the wire, even with a reduced ansatz. 
    The VQE results show the same trend as seen in previous smaller binary wires.
   }
  \label{fig:7-qca-all}
\end{figure*}

The layout for a driven, 7-cell binary wire is shown in Figure \ref{fig:0-6}. To model this circuit we used the simple, single-parameter ansatz of Figure \ref{fig:7-ansatz}. Simulated and actual hardware VQE results for this ansatz are shown in Figure \ref{fig:7-qca-all}. Even with a single parameter and the default setting for the number of shots, the VQE performed on hardware successfully models an input bit that is copied down the line of cells. 

\subsubsection{15-Cell Binary Wire}

\begin{figure}[htbp]
    \centering
    \subfloat[]{%
      \includegraphics[width=0.49\textwidth]{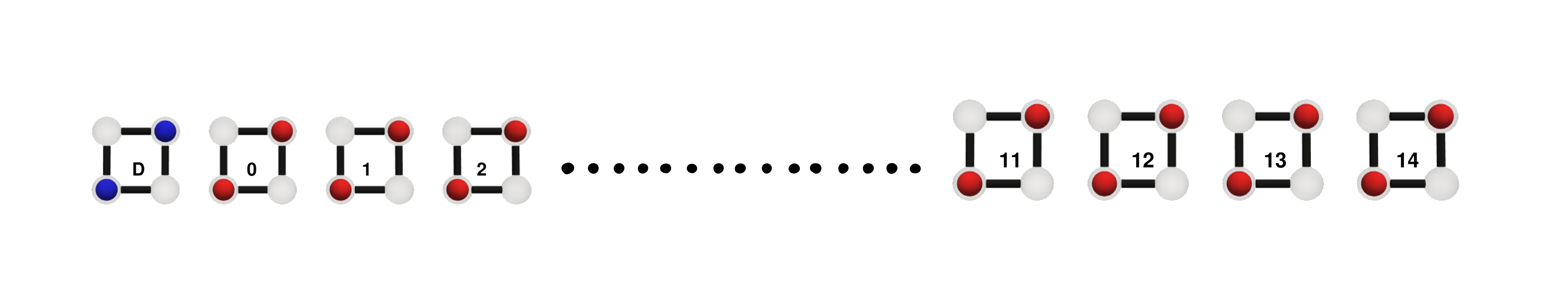}%
      \label{fig:0-14}%
    }
    \vfil
    \subfloat[]{%
      \includegraphics[width=0.45\textwidth]{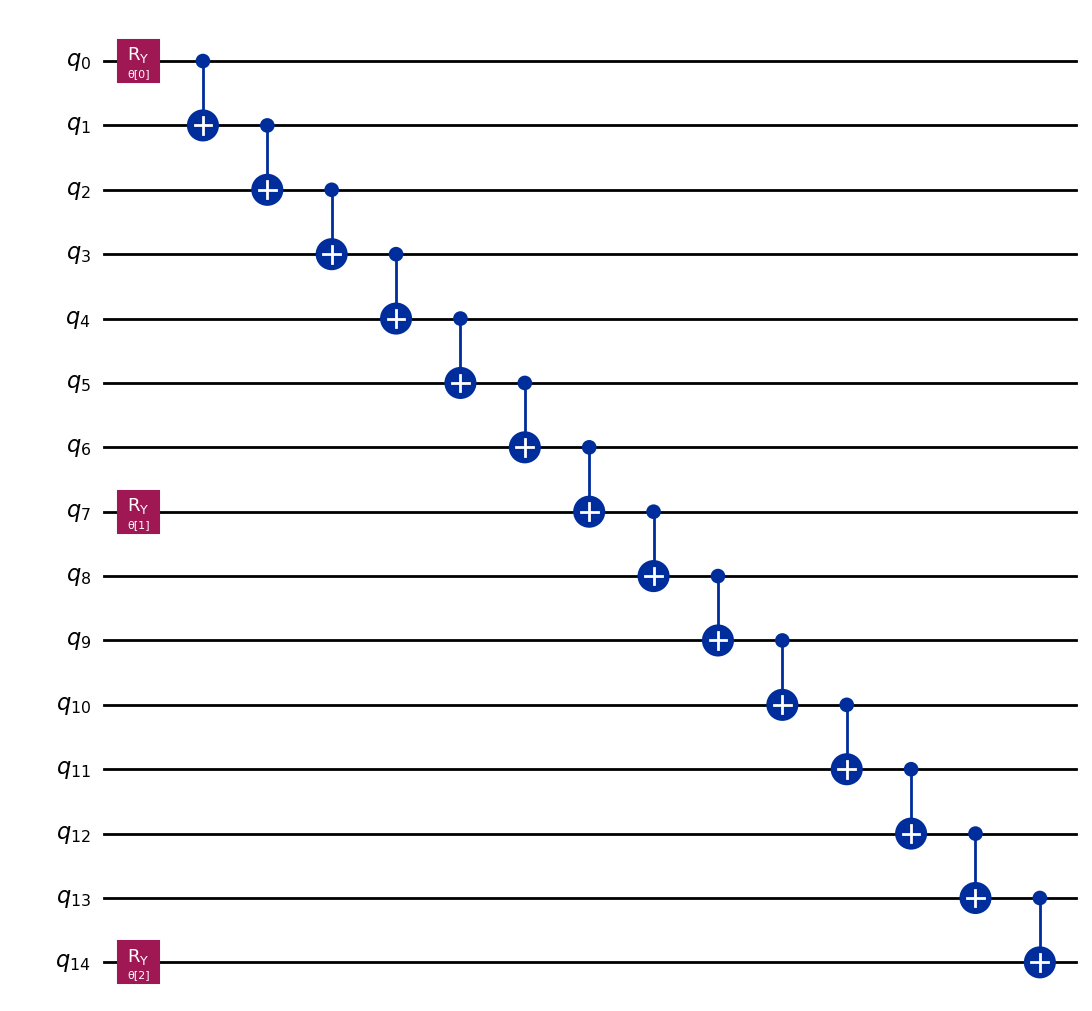}%
      \label{fig:15-ansatz}%
    }
    \caption{A fifteen-cell binary needs more parameters than previous smaller binary wires which increases the complexity of the circuit and the optimization routine. (a) An input bit drives a fifteen-cell binary wire. 
      (b) Ansatz of a fifteen cell binary wire has three rotation gates instead of just one. 
      The optimization routine is thus about three times longer and expensive than the previous VQE runs.}
    \label{fig:15-cell-layout}
\end{figure}

\begin{figure*}[htbp]
  \centering
  \includegraphics[width=\textwidth]{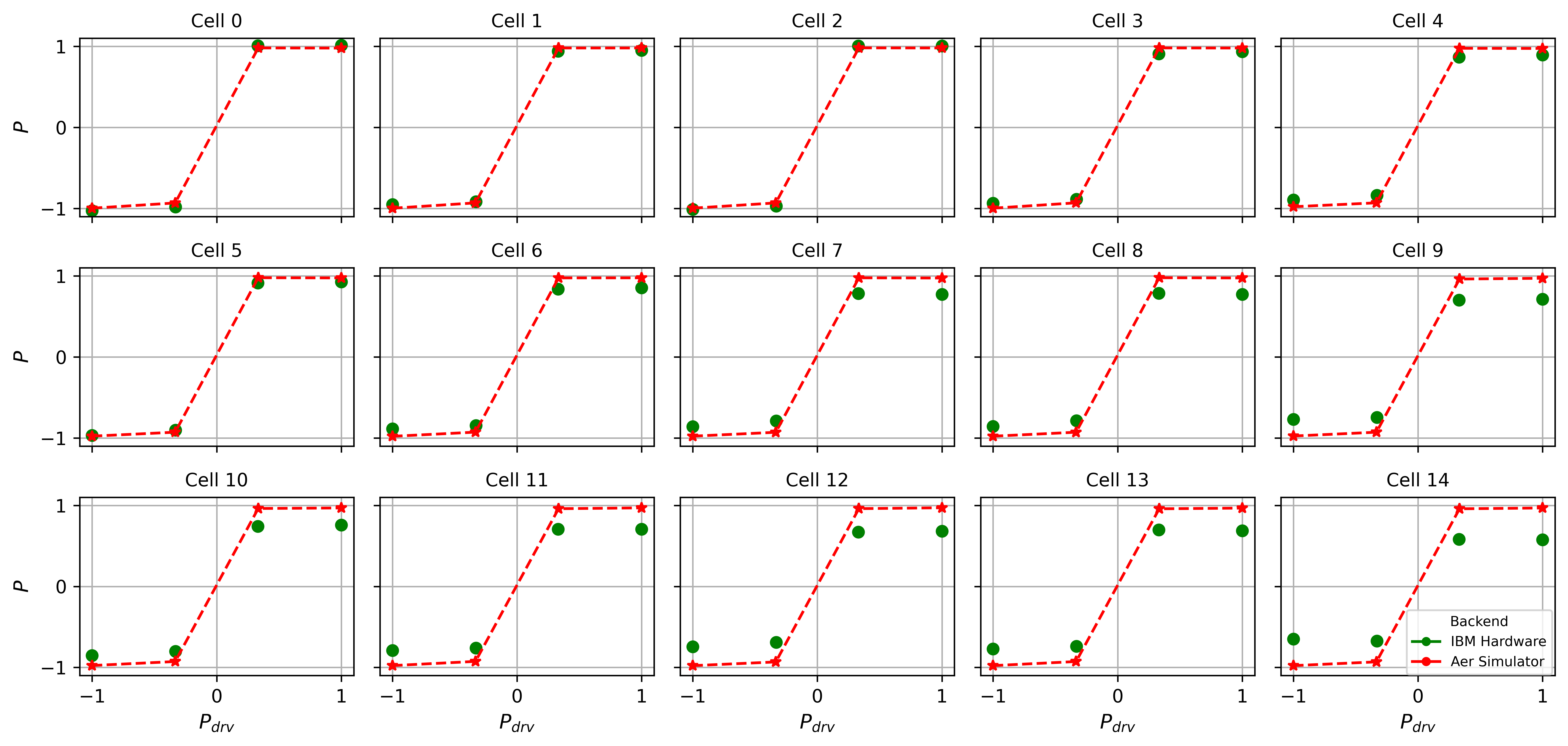}
  \caption{A similar trend as in the smaller binary wires confirms the VQE effectiveness for larger circuits. However, a slight attenuation in polarization is observed in the cells toward the end of the wire, likely due to cumulative gate noise and decoherence effects in the quantum hardware.
    }
  \label{fig:15-qca-all}
\end{figure*}

To assess the scalability and general applicability of this approach, we now extend our simulation to model the driven 15-cell binary wire, illustrated in Figure \ref{fig:0-14}. We used the ansatz shown in Figure \ref{fig:15-ansatz}, which includes three parameters, unlike parametric circuits used for smaller wires. It was necessary to include additional parametric \(\mathbf{R}_y(\theta)\) gates, since circuits with fewer \(\mathbf{R}_y(\theta)\) gates failed to model the proper binary wire response. 

The results of this extended simulation for the 15-cell QCA wire are presented in Figure \ref{fig:15-qca-all}. While Aer Simulation results were obtained, an exact, classical model of the 15-cell wire was not performed for the 15-cell case, since such a model at this scale is intractable without more involved acceleration techniques or approximations. 
The three-parameter ansatz successfully captures the expected binary wire behavior, where a driving bit is copied along the line, as seen in shorter wires. Notably, VQE enables a model of a 15-cell binary wire, whereas a fully-coherent, classical calculation is computationally prohibitive.

\begin{figure*}[htbp]
  \centering
  \includegraphics[width=\textwidth]{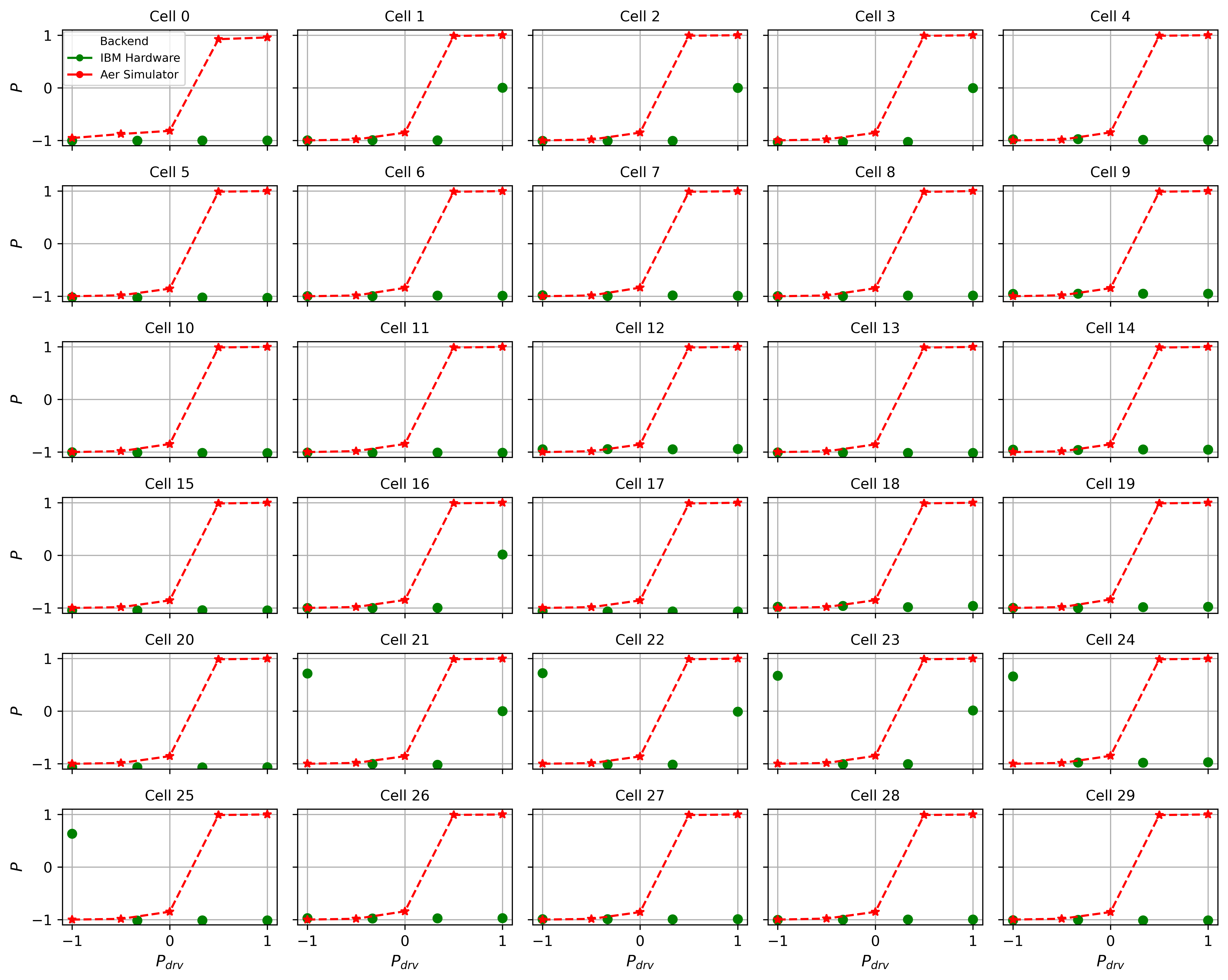}
  \caption{Noise prevented the IBM hardware from successfully modeling a thirty-cell binary wire using VQE. Noise-free VQE simulations, on the other hand, reflected a bit copy along the entire wire. }
  \label{fig:30-cell-all11}
\end{figure*}

We attempted to extend the VQE model to a 30-cell binary QCA wire, and results are shown in Figure \ref{fig:30-cell-all11}. 
While the noiseless AerSimulator backend continued to perform well, producing a polarization profile that closely matched the expected bistable behavior, the results obtained from the IBM Hardware backend were significantly degraded. 
Nonetheless, the AerSimulator simulator results affirm the potential utility for large-scale VQE QCA models once suitable quantum hardware becomes available.

\subsection{Inverter}

\begin{figure}[htbp]
    \centering
    \subfloat[]{%
      \includegraphics[width=0.3\textwidth]{img/inverter-1-0.png}%
      \label{fig:inv-layout}%
    }
    \vfil
    \subfloat[]{%
      \includegraphics[width=0.3\textwidth]{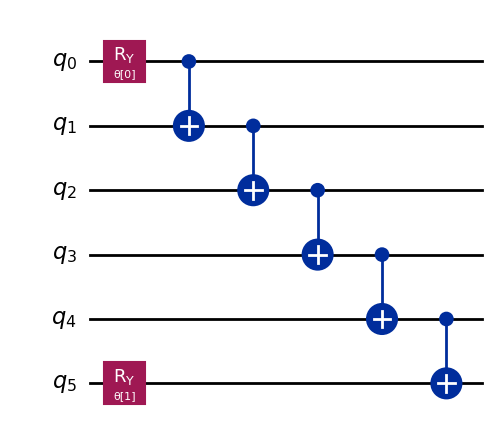}%
      \label{fig:6-inv-ansatz}%
    }
    \caption{An inverter circuit uses diagonal coupling for bit inversion and an extra rotation gate to ensure proper freedom for the last qubit to invert. (a) The final cell in the inverter circuit (cell 5) achieves logic inversion by interacting 
      diagonally with its two adjacent cells (cell 3 and cell 4). 
      (b) The reduced ansatz for the six-cell inverter has two rotation gates and thus 
      two parameters for optimization. }
    \label{fig:inv-cell-layout}
\end{figure}

\begin{figure*}[htbp]
  \centering
  \includegraphics[width=\textwidth]{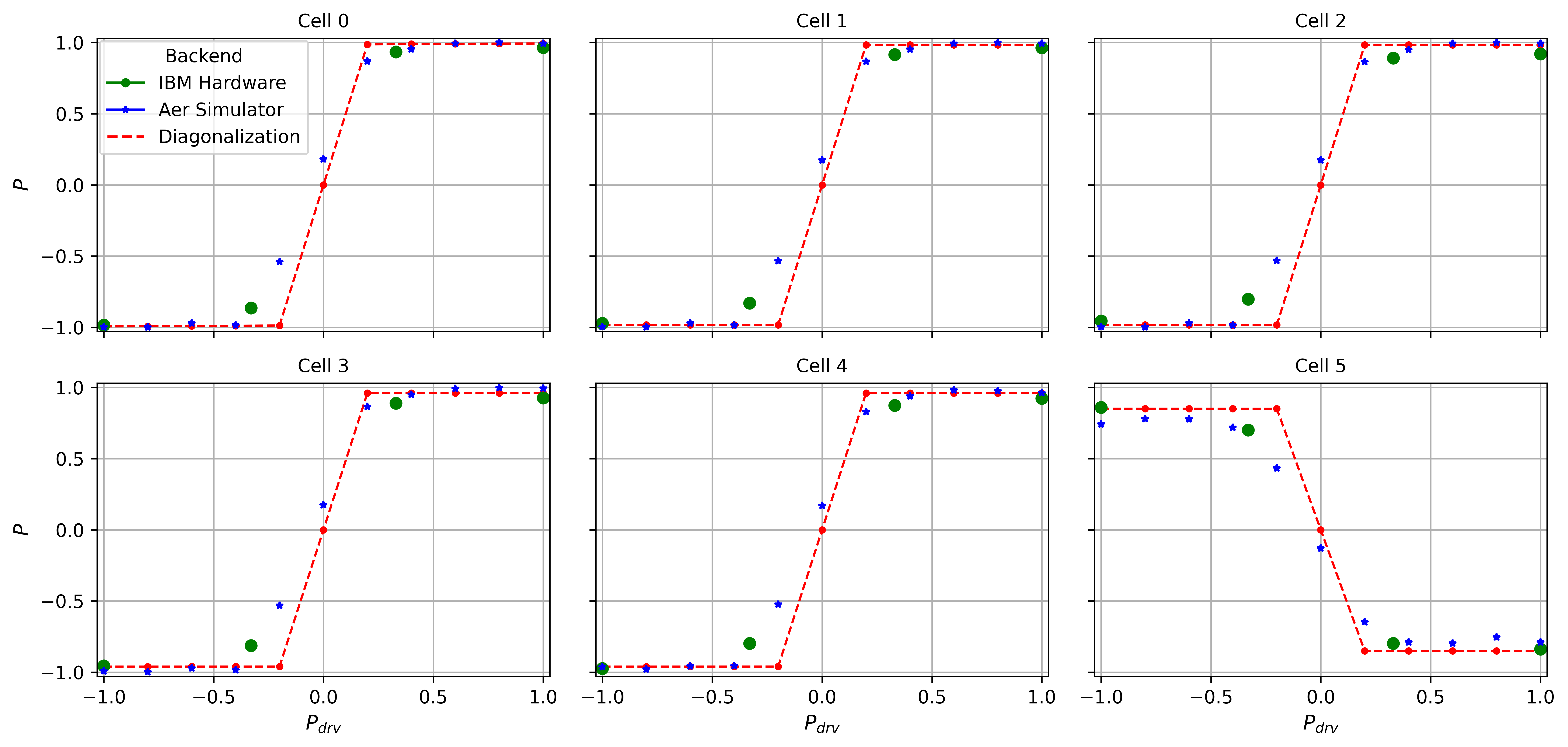}
  \caption{A VQE model of an inverter captures the desired bit inversion on cell 5 relative to 
      the driver bit, as desired. VQE results from both the noiseless 
      AerSimulator and the noisy IBM Hardware backend approximate the exact result from diagonalization, though the IBM Hardware results 
      show slightly reduced accuracy due to quantum noise and hardware-induced errors. }
  \label{fig:inv-all}
\end{figure*}

Figure \ref{fig:inv-layout} shows the layout of an inverter comprised of six cells. 
Cells 0 through 4 copy the input bit and fan it out to two bits. 

The critical operation of a bit inversion occurs at cell 5, which serves as the output of the circuit.
Unlike a nearest-neighbor interaction, cell 5 is a next-nearest neighbor to both cells 3 and 4, with a diagonal interaction that favors a bit flip. Due to this, the output logic state becomes the logical NOT of the input provided by the driver. One diagonal interaction 
could be sufficient, but the two interactions more strongly favors the bit inversion.

The ansatz depicted in Figure \ref{fig:6-inv-ansatz} has two parameters, with a \(\mathbf{R}_y(\theta_0)\) gate at the first cell and one \(\mathbf{R}_y(\theta_1)\) gate on the last qubit. The \(\mathbf{R}_y(\theta_1)\) gate on qubit 5 enables the ansatz to impose a bit flip on cell 5 relative to the other cells. With just two parameters, the ansatz circuit remains shallow. 

The VQE model of the inverter is designed to approximate the behavior of a QCA inverter, and 
the result are comparable to results from the exact calculations. The polarization response of the inverter
is shown in Figure \ref{fig:inv-all}. Our VQE model successfully captures the desired behavior: a bit \(X\) is copied from the
driver to cells 0 through 4, with an inversion to \(\bar{X}\) on cell 5. The noise-free simulation more closely approximates the
exact, classical model than do the results from the IBM Hardware, which suffer slightly more noise. Nonetheless, the VQE
model still approximates the exact classical result for the inverter ground state.

\subsection{Majority Gate}

\begin{figure}[htbp]
    \centering
    \subfloat[]{%
      \includegraphics[width=0.3\textwidth]{img/majority111.png}%
      \label{fig:6-maj-layout}%
    }
    \vfil
    \subfloat[]{%
      \includegraphics[width=0.3\textwidth]{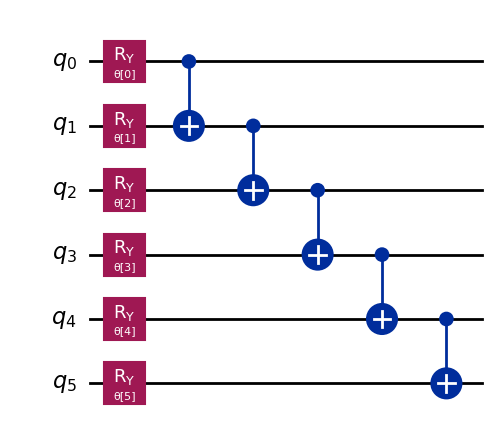}%
      \label{fig:6-maj-ansatz}%
    }
    \caption{The majority gate has the highest optimization and running cost as each cell needs freedom to interact with all the qubits and so the number of parameters is equal to the number of qubits. (a) A six-cell majority gate has three 
      drivers A, B, and C, which influence intermediate cells (0, 1, and 2). Based on the majority of these cells, cell 3 determines the output logic state, which is then propagated through cells 4 
      and 5.  
      (b) A full ansatz is needed for the majority gate and so it has
      six parameters for optimization.
       }
    \label{fig:6-cell-maj-layout}
\end{figure}

\begin{figure}[htbp]
  \centering
  \includegraphics[width=0.45\textwidth]{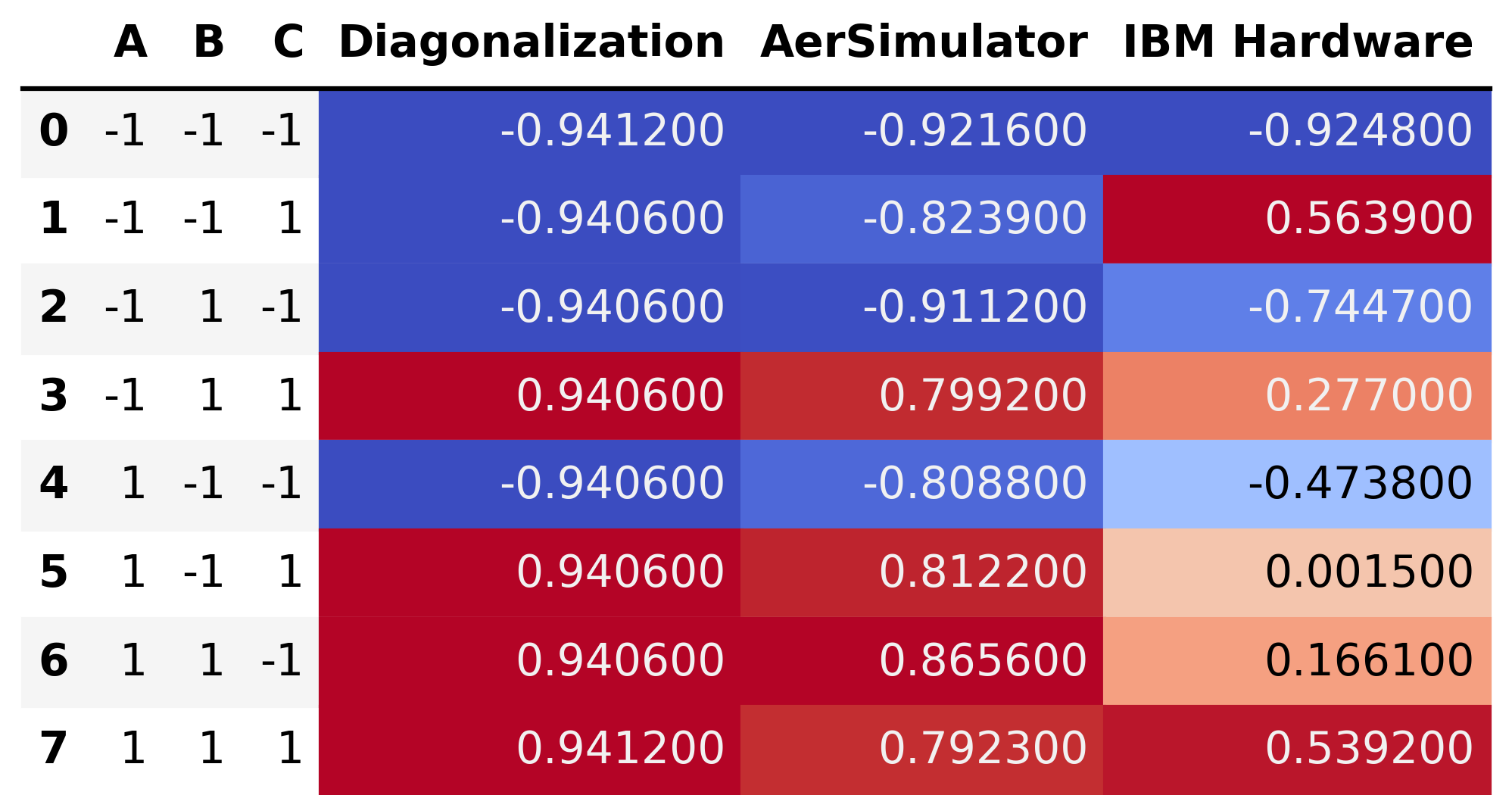}
  \caption{A VQE model of a majority gate presents a close agreement 
    between exact and noiseless simulation-based results, but the IBM Hardware fails to fully match the expected behaviour. This can be attributed to an increased number of parameters and unwanted diagonal influences from other cells.
    }
  \label{fig:majority-g}
\end{figure}

\begin{figure*}[htbp]
    \centering
    
    \includegraphics[width=\textwidth]{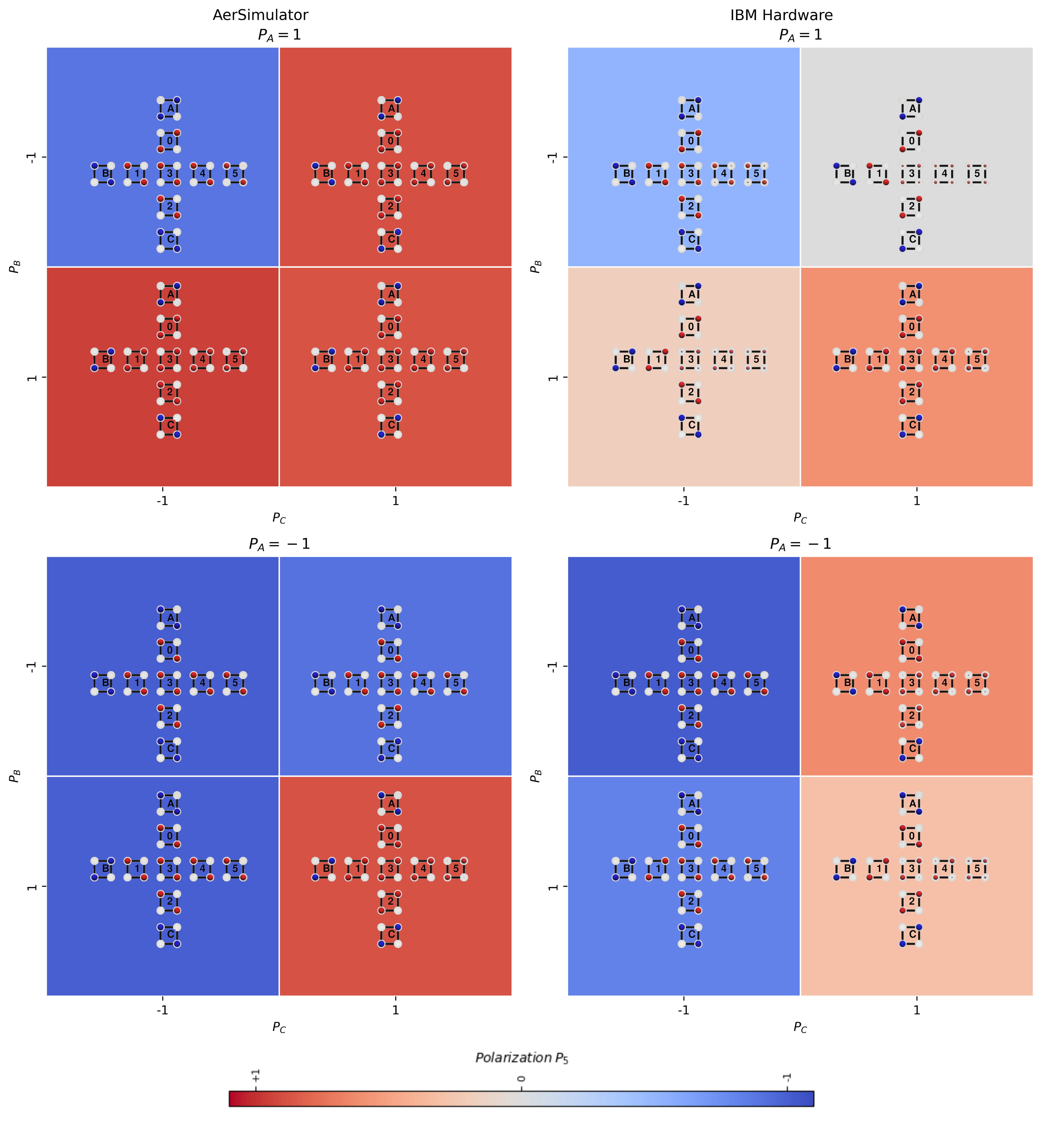}%
    \caption{The insets in the Karnaugh map help in visualizing the unintended diagonal interactions among QCA cells.
    Notably, cell 4's behavior can be seen to deviate from majority logic due to diagonal influence from cells 
    0, and 2. 
    }
    \label{fig:majority-g-coolwarm}
\end{figure*}

Figure \ref{fig:6-maj-layout} shows three driver cells (A, B, and C), and six cells (0 to 5) configured as a majority gate.
Cells 0, 1, and 2 couple the input bits to cell 3, the majority cell. Cells 4 and 5 copy the result from cell 3.

The ansatz used for the 6-cell system is presented in figure \ref{fig:6-maj-ansatz}.
Six independent parameters, \((\theta_0, \theta_1, \ldots, \theta_5\) , characterize
\(\mathbf{R}_y(\theta_k)\) gates and provide the flexibility prepare a trial wave function that represents the actual circuit ground state.
Simulations (not shown) revealed that omitting parameterized rotation gates from any qubit prevents the 
ansatz from preparing a suitable trial wavefunction.

The results of the VQE simulation of the majority gate are benchmarked against the results of diagonalization of \(\hat{H}\) and shown
in Figure \ref{fig:majority-g}.
Here, the polarization of the output cell, \(P_5\)  is listed for each of the eight fully-polarized input combinations of \(P_A = \pm 1\), \(P_B  = \pm 1\), and \(P_C = \pm 1\). For
ease of interpretation, the sign of \(P_5\) is coded in the background color of its table cell, and its magnitude maps to the color intensity
(i.e., \(P_5 < 0 \rightarrow \mbox{blue}\), and \(P_5 > 0 \rightarrow \mbox{red}\), and larger \(P_5\) corresponds to a more intense background color).
In all cases, the noise-free AerSimulator yielded the correct sign, though simulated \(P_5\) magnitudes are slightly
lower than exact magnitudes. The VQE results for \(P_5\) on quantum hardware, on the other hand, were further from the exact values.
Contributing factors are likely noise from the six parametrized gates and the five entangling gates used in the ansatz, as well as 
competing influences on cell 4 from cells 0, 2, and 3.

The competing influences are more readily apparent upon inspection of the VQE result for the entire circuit. This is visualized in
Figure \ref{fig:majority-g-coolwarm} as Karnaugh map. Again, all input combinations are shown in two sections. An upper section shows the \(A = 1\) case (\(P_A = 1\), and the lower section shows inputs where \(A = 0\) case (\(P_A = -1\). The resulting output polarization is color-coded in the background of each panel, and in the foreground, we illustrate the mean-field charge configuration for the majority gate. Results are shown for both the Aer Simulator and the quantum hardware.

We focus on the cases where the hardware result deviates the most from the noise-free VQE simulation and the exact diagonalization result. First, consider \((P_A, P_B, P_C) = (1, -1, 1)\) (the upper right panel in the top section). This is a difficult case, as is  \((P_A, P_B, P_C) = (-1, 1, -1)\), in which \(P_A = P_C = - P_B\). Here, the two bit copies \(A\) on cells 0 and 2 have a next-nearest-neighbor interaction with bit 4, biasing it toward \(\bar{A}\).  This is in competition with the nearest-neighbor interaction between bits 3 and 4: the majority bit, \(A\) is favored on cell 3, and this biases cell 4 toward \(A\). In these difficult cases, the separation between the ground state energy of the QCA circuit and its first excited state is reduced. In the VQE model, this may mean that the minimum in potential energy space over the parameter vector \(\vec{\theta}\) is rather shallow and difficult to find, especially given noise introduced in state preparation by the ansatz or noise in measurement.

Another incorrect result from the VQE on the hardware is the case where \((P_A, P_B, P_C) = (-1, -1, 1)\). This is not one of the difficult, conflicting cases. Nonetheless, since the noise-free VQE simulation found the proper majority output, it is likely that gate noise and measurement noise drove the wrong circuit result in the VQE on actual hardware.

\subsubsection{2-Cell Majority Gate}

\begin{figure}[htbp]
    \centering
    \subfloat[]{%
      \includegraphics[width=0.2\textwidth]{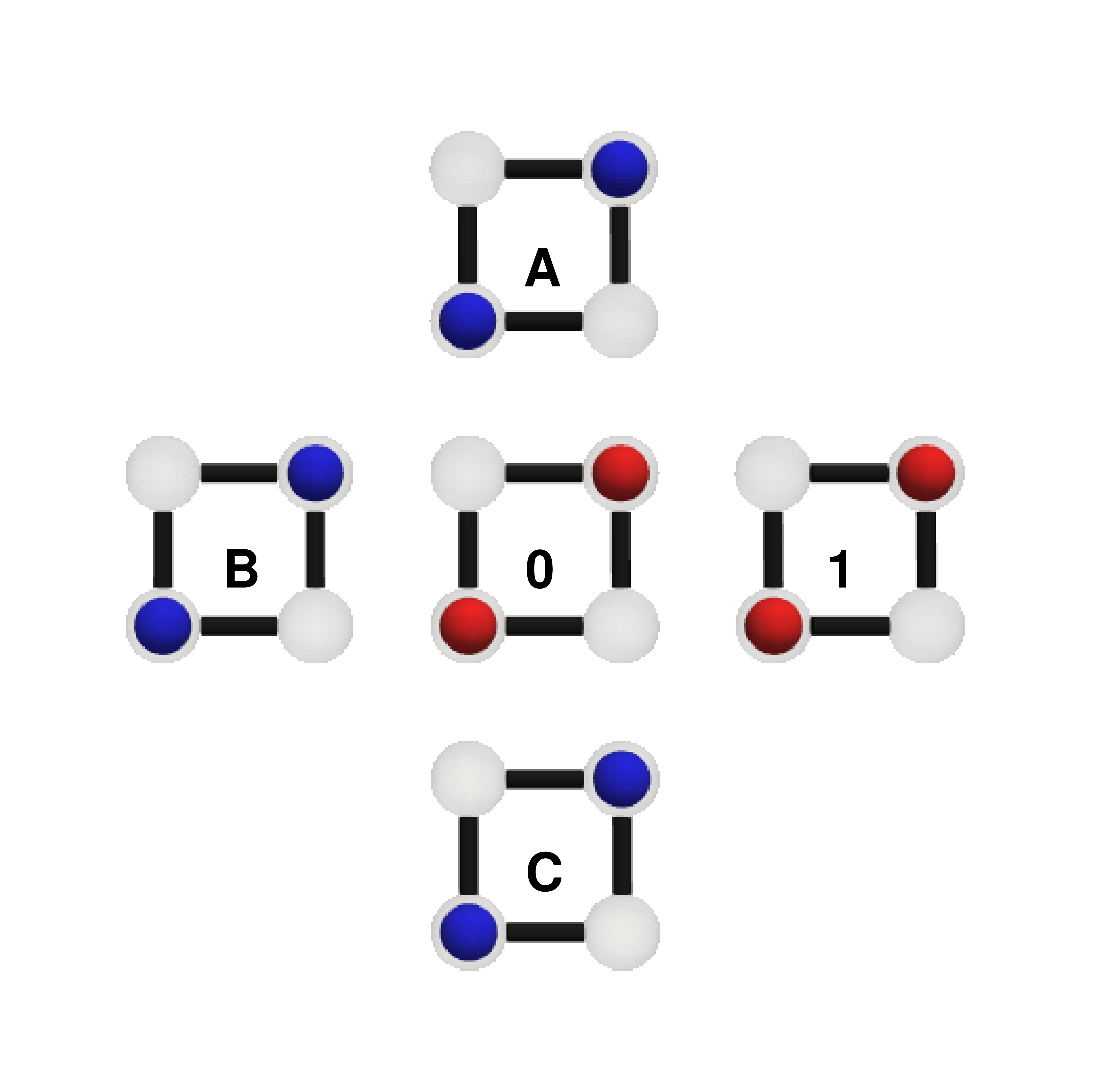}%
      \label{fig:2-maj-layout}%
    }
    \subfloat[]{%
      \includegraphics[width=0.15\textwidth]{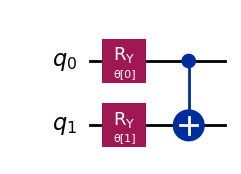}%
      \label{fig:2-maj-ansatz}%
    }
    \caption{The compact design of the two-cell majority gate reduces gate complexity and may reduce noise-related errors. (a) The layout of a two-cell majority gate is similar to the six-cell one with the connecting cells removed, keeping only the driver cells, the decision cell, and the output cell.
      (b) The ansatz is simpler than the six-cell majority gate ansatz, with only one CNOT gate and two parameterized gates.  }
    \label{fig:2-cell-maj-layout}
\end{figure}

\begin{figure}[htbp]
  \centering
  \includegraphics[width=0.45\textwidth]{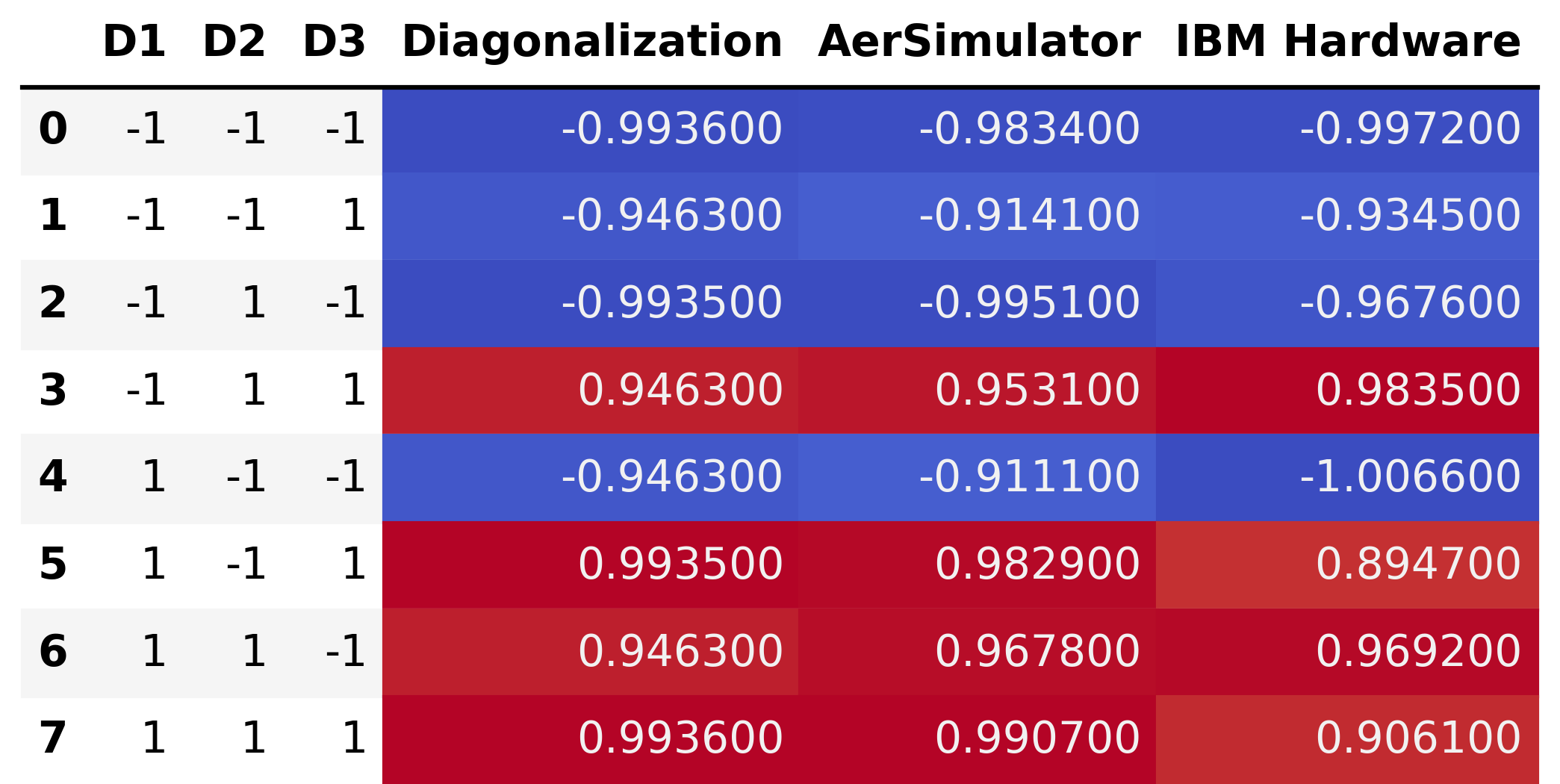}
  \caption{A VQE model of a two-cell majority gate produces the correct results for all
    eight fully polarized input combinations. This demonstrates that the two-cell majority gate VQE model is more robust than the six-cell majority gate model.  }
  \label{fig:majority-2}
\end{figure}

A simpler inverter model is easier to implement in VQE. This model reduces the QCA circuit to only
a pair of cells, as shown in Figure \ref{fig:2-maj-layout}. Here, three driver cells provide direct,
classical inputs to cell 0, the device cell. This bit is then copied to the output cell, cell 1.

To mitigate the errors in 6-cell majority gate, we introduce a streamlined 2-cell majority-gate variant 
that preserves the logical layout with three fixed driver cells setting the inputs but routes the 
signal through only two data cells. The reduced qubit count shortens the 
CNOT ladder and lowers the total exposure to decay and error accumulation and removes the unwanted 
diagonal interaction as well.
With fewer parameters to optimize, a shallower depth and no unwanted interaction, this minimalist design should 
yield markedly cleaner polarization results on NISQ hardware.

This configuration, shown in figure \ref{fig:2-maj-layout} is a compact version of the traditional majority 
gate. In this layout, three driver cells labeled A, B, and C provide the input polarizations.
However in this case, these inputs directly influence two adjacent QCA cells labeled 0 and 1, where cell 1 is the output cell. 
The ansatz designed for the 2-cell majority gate, shown in figure \ref{fig:2-maj-ansatz}, is notably 
more efficient and compact compared to the one used for the full 6-cell majority gate. It features just 
two parameterized gates. 
The simulation results are shown in 
figure \ref{fig:majority-2}. This truth table shows a really strong agreement between AerSimulator results and the IBM Hardware. 

\section{Discussion}

We have explored VQE approaches to modeling QCA circuits. To validate models and test VQE, simulated VQE and VQE models on quantum hardware were benchmarked against exact, classical results from diagonalization of the Hamiltonian whenever possible. Within a two-state approximation, the Hamiltonian--when interactions beyond the simplified next-nearest-neighbor interaction are neglected--for a QCA circuit has a simple form that lends itself to ground state energy estimations using projective measurements on quantum computers. Additionally, from the optimized circuit, we may estimate the classical polarization of cells within the approximate ground state. The VQE methods used here have enabled models of QCA circuits larger than may be treated using naive, unaccelerated, classical, fully-coherent and exact calculation of the ground state in Python; nonetheless, VQE models are affected and limited by noise, the scarcity of quantum resources, and the classical optimizer.

Noise degrades the accuracy of trial preparation and measurement, leading to some deviations from the exact results, including some unphysical results for cell polarizations. As QCA circuit sizes grow, the number of optimization parameters grows. Given the scarcity of quantum resources, it is important to minimize the number of optimization parameters so that runtime on hardware may be conserved by minimizing the number of algorithm iterations. With intuition about circuit operations, the number of parameters may be reduced. For example when bit copies are present, as in a binary wire, an \(\mathbf{R}_y\) operation may be omitted from a qubit representing a bit copy at the cost of a slight degradation of accuracy. For shorter wires up to \(N = 7\), even a single-parameter ansatz is suitable. As the length of the binary wire increases additional \(\mathbf{R}_y\) gates are required to prevent poor results on hardware. VQE models with a total of three \(\mathbf{R}_y\) gates could capture binary wire behaviors up to lengths \(N=15\) on available IBM hardware.

Beyond the binary wire, VQE models of the inverter and the majority gate were explored. Circuits exhibiting these functions may be formed using smaller networks of fewer than ten cells. In these circuits next-nearest neighbor interactions come into play. These drive anti-alignment and compete with nearest-neighbor interactions. This likely reduces the energy of excited states, making it more challenging for noisy VQE methods to arrive at the ground state.

\section{Conclusion}

VQE methods may be used to model the ground state of QCA circuits. The methods explored here used a single qubit to represent the state of a single QCA cell in an \(N\)-cell circuit within a two-state approximation. However, the performance of these methods is susceptible to noise. This may be mitigated by increasing the number of shots for higher accuracy, or using more advanced, robust, and less-noisy hardware. Additionally, it is helpful to minimize the number of parameters, which also minimizes the number of parametrized single-qubit rotation gates used. As circuit sizes grows, the number of parameters grows inevitably, resulting in increasing numbers of iterations in the classical optization. Through knowledge of circuit operation, a modeler may selectively remove gates from the ansatza at the expense of restricting  some degrees of freedom in preparing  the trial wave function, \(\ket{\Psi(\vec{\theta}}\).

While the VQE methods explored here remain very sensitive to noise, the continual development of hardware and the approach could lead to improved VQE models for QCA. Additionally, the advent of a fault-tolerant era for quantum computing could provide opportunities for leveraging methods such as quantum phase estimation (QPE) in the modeling of QCA circuit ground states. 

\begin{acknowledgments}
We acknowledge the use of IBM Quantum services for this work. The views expressed are those of the authors, and do not reflect the official policy or position of IBM or the IBM Quantum team. 
The authors also thank Colin Burdine for stimulating discussion, and for sharing IBM Quantum credits. 

\end{acknowledgments}

\vfil

\section*{Data Availability Statement}
The data that support the findings of this study are available from the corresponding author upon reasonable request.

\section*{Conflict of Interest Statement}
The authors have no conflicts to disclose.

\bibliography{references}

\end{document}